\documentclass[onecolumn]{IEEEtran}
\usepackage[english]{babel}
\usepackage{graphicx}
\usepackage{color}
\usepackage{bm}
\usepackage{amssymb}
\usepackage[intlimits]{amsmath}
\usepackage{amsmath}

\newcommand{\be}{\begin{equation}}
\newcommand{\ee}{\end{equation}}
\newcommand{\ben}{\begin{equation*}}
\newcommand{\een}{\end{equation*}}

\title{Optimal input signal distribution and per-sample
mutual information   for nondispersive nonlinear optical fiber channel at large SNR}

\author{
\IEEEauthorblockN{I.~S.~Terekhov\IEEEauthorrefmark{2}\IEEEauthorrefmark{1},
A.~V.~Reznichenko\IEEEauthorrefmark{2}\IEEEauthorrefmark{1},
Ya.~A.~Kharkov\IEEEauthorrefmark{3},  S.~K.~Turitsyn\IEEEauthorrefmark{4}\IEEEauthorrefmark{2}}\\
    \IEEEauthorblockA{\IEEEauthorrefmark{2}Novosibirsk State University, Novosibirsk, 630090, Russia}\\
    \IEEEauthorblockA{\IEEEauthorrefmark{1}Budker Institute of  Nuclear
Physics, Russian  Academy of Sciences, Novosibirsk, 630090, Russia}\\
    \IEEEauthorblockA{\IEEEauthorrefmark{3}School of Physics, University of New South Wales, Sydney 2052, Australia}\\
    \IEEEauthorblockA{\IEEEauthorrefmark{4}Aston Institute of Photonics
Technologies, Aston University, Aston Triangle, Birmingham, B4 7ET, UK}
}

\begin{document}
\maketitle

%\hskip 2cm
%\date{\today}

\begin{abstract}

We consider a model nondispersive nonlinear optical fiber channel with additive
white Gaussian noise at large $\mathrm{SNR}$ (signal-to-noise ratio) in the
intermediate power region. Using Feynman path-integral technique we for the first
time find the optimal input signal distribution maximizing the channel's per-sample
mutual information. The finding of the optimal input signal distribution
allows us to improve previously known estimates for the channel capacity. The output
signal entropy, conditional entropy, and per-sample mutual information are calculated for Gaussian,
half-Gaussian and modified Gaussian input signal distributions. We explicitly show
that in the intermediate power regime the per-sample mutual information for the
optimal input signal distribution is greater than the per-sample mutual information
for the Gaussian and half-Gaussian input signal distributions.

\end{abstract}

%\keywords{conditional entropy,  mutual information, nonlinear Shr\"{o}dinger
%equation.} \pacs{05.10.Gg, 89.70.-a,  02.70.-c,02.70.Rr,05.90.+m}

%==================Introduction
\section{Introduction.}

The channel capacity $C$  introduced by Shannon in his seminal work
\cite{Shannon:1948} is related to the maximum amount of information that can be
reliably transmitted over a noisy communication channel.  Shannon calculated the capacity of the linear channel
with additive white Gaussian noise (AWGN) and found the famous logarithmic
dependence of the channel's capacity on the signal power:
\begin{eqnarray}\label{CapacityShannon}
C\propto\log_{2}\left(1+\mathrm{SNR}\right)\,,
\end{eqnarray}
where $\mathrm{SNR}=P/N$ is the signal-to-noise power ratio, $P$ is the signal
power, and $N$ is the noise power. This, in particular, means that when the noise
power $N$ is fixed, in order to increase the capacity one has to increase the signal
power $P$.

The interest in nonlinear communication channels has been increasing since the beginning
of the 2000's when fiber optics communication systems had to increase both bandwidth
and system reach which required the use of ever higher optical power. Fiber optic
nonlinear channels have been studied both analytically and
numerically in numerous papers, see e.g.
\cite{Mitra:2001,Narimanov:2002,Kahn:2004,Essiambre:2008,Essiambre:2010,
Killey:2011,Agrell:2014,Sorokina2014} and references therein. The simplified model nondispersive
nonlinear optical fiber  channel  was considered, e.g. in
\cite{M1994,MS:2001,Tang:2001,tdyt03,Mansoor:2011}. The investigation of nonlinear
communication channels where transmission is affected and changed by the signal
power is a difficult problem, especially at large $\mathrm{SNR}$
\cite{Essiambre:2010}. Analysis of the capacity of these channels is technically
challenging and new techniques and methods are highly desirable to advance these
studies \cite{Narimanov:2002,tdyt03,Agrell:2012,Agrell:2013,Terekhov:2014}.
In this work we consider a simplified model nonlinear channel with a limited range of practical applications.
However, methods developed for and tested on such model channels might be
useful for much more complex and challenging nonlinear fiber communication problems. We introduce here a new approach to the calculation of the conditional probability density function via the path-integral technique
and demonstrate its application using considered model channel as a particular example.

%===========================NOTIONS
The channel capacity $C$ is defined as the maximum of the mutual information $I_{P_{X}[X]}$
with respect to the probability density function $P_{X}[X]$ of the input signal $X$:
\begin{eqnarray}
C=\max_{P_{X}[X]} I_{P_{X}[X]} ,\label{capacity1}
\end{eqnarray}
where the maximum value of $I_{P_{X}[X]}$ should be found subject to the condition
of fixed average signal power:
\begin{eqnarray} \label{power}
P=\int {\cal D} X |X|^2 P_{X}[X].
\end{eqnarray}
The mutual information of a memoryless channel is defined in terms of the output
signal  entropy $H[Y]$  and conditional entropy $H[Y|X]$:
\begin{eqnarray}
& I_{P_{X}[X]}= H[Y]-H[Y|X], \label{MI}
\end{eqnarray}
with
\begin{eqnarray}
 \!\!\! H[Y|X]&=&-\int {\cal D}X {\cal D}Y P_{X}[X]{P[Y| X]} \log P[Y| X],
\label{condentropy}\\
\!\!\! H[Y]&=&-\int{\cal D}Y P_{out}[Y] \log P_{out}[Y],
\label{entropies}\\
\!\!\! P_{out}[Y]&=&\int {\cal D}X P_{X}[X] P[Y|X], \label{Pout}
\end{eqnarray}
where $P[Y|X]$ is the conditional probability density function (PDF) for an output signal
$Y$ when the input signal is $X$, and $P_{out}[Y]$ is the PDF for an output signal $Y$. The measure ${\cal D}Y$ is defined as $\int {\cal
D}Y P[Y|X] =1$, and ${\cal D}X$ is defined as $\int {\cal D}X P_{X}[X] =1$. The
capacity (\ref{capacity1}), as defined by (\ref{MI})-(\ref{Pout}), is measured in
units of $(\log 2)^{-1}$ bits per symbol (also known as nats per symbol). The input
and output signals are functions of time where the signal's spectrum is restricted to a given
bandwidth. In general, a sampling of the temporal signal should be introduced to
define a discrete-time memoryless channel, however, here we consider only per-sample
quantities.

The channel's mutual information (\ref{MI}) depends on the probability distribution
$P_X[X]$  of the input signal. The input signal PDF, that maximizes the channel's
per-sample  mutual information is called ``capacity-approaching'' or ``optimal''
PDF $P_X^{opt}[X]$. Obviously, the problem of finding the optimal PDF of the input
signal for nonlinear optical channels is of great practical importance.

In the previous studies of nondispersive nonlinear optical channels (e.g. \cite{MS:2001}, \cite{tdyt03},  \cite{Mansoor:2011}) the Gaussian and half-Gaussian input signal PDF's were used as trial functions in order to put low bound constraint on the channel capacity, or to provide asymptotic estimate of the capacity in the regime of large SNR. The authors of \cite{Mansoor:2011} argued, that the half-Gaussian PDF which we denote as $P_X^{(1)}[X]$,
\begin{eqnarray} \label{halfgauss}
P_X^{(1)}[X]=\dfrac{\exp\left\{-|X|^2/(2 {P})\right\}}{\pi |X| (2 \pi { P} )^{1/2}},\,
\end{eqnarray}
provides the best approximation for the ``capacity-approaching'' input signal
distribution at large SNR. In the present paper by solving a variational problem we
show that it  is not the case.  We find a true optimal distribution $P_X^{opt}[X]$
(which in fact is different from half-Gaussian distribution) in the regime of large
SNR for intermediate power range. We explicitly show, that in this regime the mutual
information (\ref{MI}) for our optimal input signal PDF is larger than the mutual
information for the Gaussian and half-Gaussian input signal distributions.

The estimates for the capacity of nonlinear fiber channels with zero dispersion and
additive white Gaussian noise in the regime of large SNR were obtained in Refs.
\cite{tdyt03},  \cite{Mansoor:2011}.
%
%Martin-Siggia-Rose formalism based on quantum field theory methods \cite
%{Zinn-Justin}.
%The analytical expression for the conditional probability density function of the channel was
%obtained in the form of an infinite series \cite{tdyt03,M1994} within the
%Martin-Siggia-Rose formalism based on quantum field theory methods \cite
%{Zinn-Justin}.
The lower bound  for capacity of the channel, based on trial Gaussian input signal
PDF, reads~\cite{tdyt03}:
\begin{eqnarray}
\label{Capacityboundary} C \geq \frac{1}{2}\log\left(\mathrm{SNR}\right)+
\frac{1+\gamma_{E} -\log (4 \pi )}{2}+{\cal O}\left(\frac{\log\left(
\mathrm{SNR}\right)}{ \mathrm{SNR}}\right)\,,
\end{eqnarray}
where $\gamma_{E} \approx 0.5772$ is the Euler constant. Note that the second term
on the right-hand side of Eq.~(\ref{Capacityboundary}) was presented as ${\cal
O}(1)$ in Ref.~\cite{tdyt03} but it is easily calculated using Eqs. (23) and (24) of
Ref.~\cite{tdyt03}. The pre-logarithmic factor $1/2$ in Eq.~(\ref{Capacityboundary})
arises as a result of the fact that in the high power regime, when the signal power
$P \gtrsim \Big(N \gamma^2 L^2 \Big)^{-1}$, the signal-dependent phase noise due to
self phase modulation occupies the entire phase interval $[0,2\pi]$ and, as a result,
the phase does not transfer information, see Ref.~\cite{Mansoor:2011}. Here $\gamma$
is the Kerr nonlinearity coefficient and $L$ is the fiber link length, see below. In
\cite{Mansoor:2011} capacity estimates were also given in the intermediate power
range $N \ll P \ll 6 \pi ^2 \Big(N \gamma^2 L^2 \Big)^{-1}$. For such a power $P$
the following estimate of the lower bound for the capacity, based on the half-Gaussian input signal PDF, was derived
\cite{Mansoor:2011}:
\begin{eqnarray} \label{capacityyousefi}
C \geq -\log(\gamma N L) + \frac{\gamma_{E}-1+\log (3 \pi )}{2}+  {\cal
O}\left(\frac{1}{\sqrt{\mathrm{SNR}}}\right),
\end{eqnarray}
where instead of ${\cal O}\left(\frac{1}{\sqrt{\mathrm{SNR}}}\right)$ the authors
presented the explicit function of the parameter $\mathrm{SNR}$ which decreases at
large $\mathrm{SNR}$, see Eq.~(40) in \cite{Mansoor:2011}.   However, the authors of
\cite{Mansoor:2011} did not take into account the ${1}/{\sqrt{\mathrm{SNR}}}$
corrections in the output signal entropy $H[Y]$, therefore, using these explicit
functions in the capacity inequality is beyond the calculation accuracy. It also
means that the result Eq.~(40) of \cite{Mansoor:2011} is not a lower bound on the
capacity. It is worth noting that in their result there is term $\log 2$ missing.
Also their result does not recover the Shannon limit $\log\, \mathrm{SNR}$ as
$\gamma \to 0$. Moreover, it is strange that the capacity estimate goes to infinity
when $\gamma$ tends to zero.
Therefore, there are obvious flaws in the result (\ref{capacityyousefi}).
% Therefore this result is obviously incorrect.

%To obtain
%the estimate (\ref{capacityyousefi}) the half-Gaussian {distribution} $P_X^{(1)}[X]$
%of the input signal was exploited as the capacity-approaching distribution:
%\begin{eqnarray} \label{halfgauss}
%P_X^{(1)}[X]=\dfrac{\exp[-|X|^2/(2 {P})]}{\pi |X| (2 \pi { P} )^{1/2}},\,
%\end{eqnarray}
%where ${P}$ is the average power of the input signal.

The analytical expression for the conditional probability density function of the
channel was obtained in the complex form of an infinite series
\cite{M1994,tdyt03,Mansoor:2011} within the Martin-Siggia-Rose formalism based on
quantum field theory methods \cite {Zinn-Justin}. In the present paper we adopt the
Martin-Siggia-Rose formalism and develop a new method for the approximate
computation of the conditional probability density function $P[Y|X]$. Using this
method we obtain the simple analytical expression for the function $P[Y|X]$ in the
leading and next-to-leading order in the parameter $1/\mathrm{SNR}$ for the
intermediate power regime
\begin{eqnarray} \label{IPR}
N \ll P \ll \Big(N \gamma^2 L^2 \Big)^{-1}.
\end{eqnarray}
Our method allows us first to derive the analytical expression for the mutual
information and then the optimal input signal distribution $P^{opt}_X[X]$ which
is different from the half-Gaussian.

%===========================MODEL and BRIEF RESULTS

In \cite{Terekhov:2014} a method to calculate the conditional PDF for a nonlinear
optical fiber channel with nonzero dispersion in the large $\mathrm{SNR}$ limit  was
introduced. Here we illustrate this general approach in application to a simpler
nondispersive nonlinear optical fiber channel as considered in
\cite{tdyt03,Mansoor:2011,M1994}. Since the channel is dispersionless, the temporal
signal waveform does not change during propagation (note, though, that the signal
bandwidth will grow due to the fiber nonlinearity and signal modulation). Therefore,
instead of considering the evolution of $\psi(z,t)$ we can consider a set of
independent scalar channels \cite{M1994,Mansoor:2011} (per-sample channels) governed
by the following model:
\begin{align}
& \partial_{z}\psi(z)-i\gamma |\psi(z)|^2 \psi(z)=\eta(z),
\label{Shrodingerequation}
\end{align}
where $\psi(z)$ is the signal function that is assumed to obey the boundary
conditions $\psi(0)=X$, $\psi(L)=Y$. The noise $\eta(z)$ has zero mean $\langle \eta
(z) \rangle_{\eta} = 0$ and a correlation function $\langle \eta (z)\bar{\eta}(z')
\rangle_{\eta} = Q \delta(z-z^\prime)\,$, so that the $\mathrm{SNR}=P/QL$, where $P$
and $N=QL$ are  the per-sample signal power and the per-sample noise power, respectively.
The connection between the differential model (\ref{Shrodingerequation}) and the
conventional information-theoretic presentation in the form of an explicit
input-output probabilistic model and appropriate sampling has been discussed in
detail in \cite{tdyt03,Mansoor:2011,M1994}. For this per-sample channel we calculate
the conditional probability density function (in order to illustrate how our method
works), the conditional entropy (\ref{condentropy}), the output signal entropy
(\ref{entropies}), and the mutual information (\ref{MI}). Solving a variational
problem for the mutual information we find the optimal input signal distribution
$P_{X}[X]$ maximizing the mutual information in the leading order in
$1/\mathrm{SNR}$.

The paper is organized as follows. In Section~\ref{section2} we develop the
quasi-classical method for the calculation of the conditional PDF $P[Y| X]$ for
arbitrary nonlinearity in the intermediate power regime (\ref{IPR}) in the leading
and next-to-leading order in $1/\mathrm{SNR}$. We find a simple representation for
$P[Y| X]$ in this case. This allows us to calculate the output signal distribution
$P_{out}[Y]$. The optimal signal distribution $P_X^{opt}[X]$ is found in Section
\ref{section3}. Section \ref{section4} is focused on the calculation and the
comparison of the mutual information for various input signal distributions. We
demonstrate that there is a range of power $P$ where the mutual information
$I_{P_X^{(2)}[X]}$ calculated for a Gaussian distribution $P_X^{(2)}[X]$, see
Eq.~(\ref{PX}) below, is closer to $I_{P_X^{opt}[X]}$, whereas at large enough power
$P$ the mutual information $I_{P_X^{(1)}[X]}$ calculated for the half-Gaussian
distribution $P_X^{(1)}[X]$ is closer to $I_{P_X^{opt}[X]}$ than the mutual
information $I_{P_X^{(2)}[X]}$. We discuss our results in Section \ref{section5}.

%==========================================Conditional probability
%\section{"Quasiclassical" method of calculation of the conditional probability in the large $\mathrm{SNR}$ limit}
\section{The conditional PDF $P[Y|X]$ and output signal PDF $P_{out}[Y]$ at large $\mathrm{SNR}$}
\label{section2}
%=================
\subsection {"Quasiclassical" method for the conditional PDF $P[Y|X]$ calculation}

The conditional probability density function can be written via the path-integral
form \cite{tdyt03,Zinn-Justin, Feynman} in a retarded discretization scheme, see
e.g. Supplemental Materials  of  Ref.~\cite{Terekhov:2014}
\begin{eqnarray} \label{ProbabInitial}
&P[Y|X] = \!\!\!\! \int\limits_{\psi(0)=X}^{\psi(L)=Y}  \!\!\!\! {\cal D}\psi
\exp\Big\{-\frac{S[\psi]}{Q}\Big\} \,,
\end{eqnarray}
and can be reduced to the quasi-classical form, see Ref.~\cite{Feynman}:
\begin{eqnarray}
\label{QuasiclassProbabInitial} P[Y|X] = e^{-\frac{S[\Psi_{cl}(z)]}{Q}} \!\!\!\!\!
\int\limits_{\tilde{\psi}(0)=0}^{\tilde{\psi}(L)=0}\!\!\!\!\! {\cal D}\tilde{\psi}
\,e^{-\frac{S[\Psi_{cl}(z)+\tilde{\psi}(z)]-S[\Psi_{cl}(z)]}{Q}}\,,
\end{eqnarray}
where the effective  action $S[\psi]= \int\limits_{0}^{L}dz
\Big|\partial_{z}\psi-i\gamma |\psi|^2 \psi\Big|^2$, and the function $\Psi_{cl}(z)$
is the "classical" solution of the equation $\delta S[\Psi_{cl}]=0$, where $\delta
S$ is the variation of our action $S[\psi]$. The equation $\delta S[\Psi_{cl}]=0$
(Euler-Lagrange equation) has the form
\begin{eqnarray}
\label{classicalTrajectoryEq} \frac{d^2\Psi_{cl}}{dz^2}-4i\gamma
\left|\Psi_{cl}\right|^2 \frac{d\Psi_{cl}}{dz}-3\gamma^2 \left|\Psi_{cl}\right|^4
\Psi_{cl}=0,
\end{eqnarray}
with the boundary conditions $\Psi_{cl}(0)=X$, $\Psi_{cl}(L)=Y$.

In order to find $P[Y|X]$ one should calculate the exponent
$e^{-\frac{S[\Psi_{cl}(z)]}{Q}}$ and the path-integral in
Eq.~(\ref{QuasiclassProbabInitial}).  First, we evaluate the exponent. To find it we
have to calculate the function $\Psi_{cl}(z)$ and then the action $S[\Psi_{cl}(z)]$.
We found the general solution $\Psi_{cl}(z)$ of (\ref{classicalTrajectoryEq})
implicitly through the boundary conditions, see
Eqs.~(\ref{Asolutiontrig})--(\ref{Aboundary}), and Eq.~(\ref{Asolutionhyp}) in
Appendix~\ref{Appendix1}. This form of the solution is inconvenient for further
calculations. Therefore we adopt a different approach and find the solution in the
leading and next-to-leading order in $1/\mathrm{SNR}$, linearizing Eq.
(\ref{classicalTrajectoryEq}) in the vicinity of the solution $\Psi_0(z)$. Here
$\Psi_0(z)$ is the solution of the equation (\ref{Shrodingerequation}) with zero
noise and with the boundary condition $\Psi_0(0)=X=\rho e^{i\phi^{(X)}}$. The
function $\Psi_0(z)$ reads
\begin{eqnarray}
\label{zeronoisesolution} \Psi_0(z)=\rho \exp\left\{i \mu \frac{z}{L}+ i
\phi^{(X)}\right\},
\end{eqnarray}
where $\mu=\gamma L \rho^2 = \gamma L |X|^2$. Note that this solution satisfies only
the input boundary condition $\Psi_0(0)=X=\rho e^{i\phi^{(X)}}$, and it is the
solution of Eq.~(\ref{classicalTrajectoryEq}) as well. Therefore, we look for the
solution of Eq. (\ref{classicalTrajectoryEq}) in the following form
\begin{eqnarray} \label{psiclasskappa}
\Psi_{cl}(z)=\Big(\rho+\varkappa(z)\Big)\exp\left\{i \mu \frac{z}{L}+i
\phi^{(X)}\right\},
\end{eqnarray}
where  the function $\varkappa(z)$ is assumed to be small: $|\varkappa(z)| \ll
\rho$. In the general case, the ratio $|\varkappa(z)|/ \rho$ is not necessarily
small  and it depends on the output boundary condition $\varkappa(L)$. However, the
configurations of $\varkappa(z)$ at which $\Psi_{cl}(z)$ significantly deviates from
$\Psi_0(z)$ ($|\varkappa(z)|\sim \rho$) are statistically irrelevant.
%KJB does this (last statement) follow from the quantum analogy?
Indeed, the expansion $S[\Psi_0(z)+\delta \Psi(z)]\propto \varkappa^2(z)$ starts
from the quadratic term at small $\varkappa(z)$, since the action achieves an
extremum (the absolute minimum $S[\Psi_0(z)] = 0$) on the solution $\Psi_0(z)$. Thus
the exponent $e^{-\frac{S[\Psi_{cl}(z)]}{Q}}$ and, as a result, the conditional PDF
$P[Y|X]$ vanishes exponentially if the typical $\varkappa(z)$ is much greater than
$\sqrt{QL}$.

Substituting Eq.~(\ref{psiclasskappa}) into Eq.~(\ref{classicalTrajectoryEq}) and
retaining only terms linear in $\varkappa(z) / \rho$, we obtain the following
equation which is still exact in the non-linearity parameter $\mu$:
\begin{eqnarray} \label{kappaequation}
\frac{d^2\varkappa}{dz^2}-2i\frac{\mu}{L}\frac{d\varkappa}{dz}-4 \frac{\mu^2}{L^2}
Re[\varkappa]=0.
\end{eqnarray}
The boundary conditions for the function $\varkappa(z)$ read
\begin{eqnarray}\label{kappaboundary}
\varkappa(0)=0,\,\varkappa(L)=Y e^{-i \phi^{(X)}-i\mu}-\rho \equiv x_0+iy_0,
\end{eqnarray}
where $x_0= Re\{ \varkappa(L)\}$ and $y_0= Im \{\varkappa(L)\}$.  The solution of
the linearized boundary problem (\ref{kappaequation}), (\ref{kappaboundary}) reads
\begin{eqnarray}\label{kappasolution}
&&Re [\varkappa(z)]=\Big(  \mu \frac{\mu x_0-y_0}{1+\mu^2/3} \frac{z}{L}+
\frac{(1-2\mu^2/3)x_0+\mu y_0}{1+\mu^2/3} \Big)\frac{z}{L}, \nonumber \\&& Im[
\varkappa(z)]=\Big( \frac{\mu x_0-y_0}{1+\mu^2/3} \left(\frac{2\mu^2
z^2}{3L^2}-1\right) + \mu\, \frac{(1-2\mu^2/3)x_0+\mu y_0}{1+\mu^2/3}
\frac{z}{L}\Big)\frac{z}{L}.
\end{eqnarray}

After substitution of the solution  Eq. (\ref{kappasolution}) in the action  we
obtain
\begin{eqnarray}\label{actionkappa}
&& \frac{1}{Q}S[\Psi_{cl}(z)]=\frac{1}{Q}S\left[\left(\rho+\varkappa(z)\right)\exp\left\{i \mu \frac{z}{L}+i
\phi^{(X)}\right\}\right] \approx \nonumber
\\&&  \frac{1}{Q}\int^{L}_0 dz \Big|\partial_z \varkappa- 2 i \frac{\mu}{L}
Re[  \varkappa]\Big|^2 = \nonumber \\&& \frac{(1+4\mu^2/3)x^2_0-2\mu x_0
y_0+y^2_0}{Q L(1+\mu^2/3)}.
\end{eqnarray}
Note that here we retain only the terms quadratic in $\varkappa$. However, it is
straightforward to calculate the next correction to the action (\ref{actionkappa})
which is  $O(1/\sqrt{\mathrm{SNR}})$, see details in Appendix \ref{Appendix1}. A
regular perturbative expansion for $\varkappa(z)$ in powers of
$1/\sqrt{\mathrm{SNR}}$ can be obtained using the exact equation for the function
$\varkappa(z)$, see  Eq.~(\ref{Akappaequationexact}) in Appendix \ref{Appendix1}.

The next step in evaluation of the conditional probability $P[Y|X]$ is the
calculation of the path-integral in Eq.~(\ref{QuasiclassProbabInitial}). In order to
calculate the path-integral in the leading $1/\mathrm{SNR}$ order we retain only
quadratic in $\tilde{\psi}$ terms in the integrand.
%The higher order terms result in the corrections to the path-integral,
%which contain an extra power of the parameter $Q L$ therefore these corrections can
%be neglected at large $\mathrm{SNR}$.
Any extra power of $\tilde\psi$ or $\varkappa$ is suppressed by the
multiplicative parameter $\sqrt{QL}$, because at small $Q$ the main contribution to
the path-integral comes from  $\tilde\psi\sim \sqrt{QL}$.
Moreover, as soon as we calculate the
path-integral in the leading order in $Q$, we can substitute $\Psi_{0}(z)$ for
$\Psi_{cl}(z)$ in the action difference
$S[\Psi_{cl}(z)+\tilde{\psi}(z)]-S[\Psi_{cl}(z)]$. To find $P[Y|X]$ in the
next-to-leading order in $1/\mathrm{SNR}$ we should retain both $\varkappa(z)$ in
$\Psi_{cl}(z)$ and higher powers of $\tilde{\psi}$ in the action difference. Details
of the path-integral calculation in the leading and next-to-leading order in
$1/\sqrt{\mathrm{SNR}}$ are presented in Appendix \ref{Appendix2}. Taking into
account the expression for the action (\ref{AactinNLO}) and the result of the
path-integral calculation (\ref{Acuantumcorrfinal}) we obtain the final result
for $P[Y|X]$:
\begin{eqnarray}\label{PYXapprox}
&&\!\!\!P[Y|X]=\dfrac{\exp\left\{- \dfrac{(1+4\mu^2/3)x^2_0-2\mu x_0 y_0+y^2_0}{Q L
(1+\mu^2/3)}\right\}} {\pi Q L \sqrt{1+\mu^2/3}}\Bigg(1- \frac{\mu/\rho}{15
(1+\mu^2/3)^2}\left(\mu (15+\mu^2)x_0-2(5-\mu^2/3)y_0\right)- \nonumber \\&&
\frac{\mu/ \rho}{135 Q L \left(1+\mu ^2/3\right)^3 }\Big\{ \mu \left(4 \mu ^4+15 \mu
^2+225\right) x^3_0+ \left(23 \mu ^4+255 \mu ^2-90\right) x^2_0 y_0+  \mu \left(20
\mu ^4+117 \mu ^2-45\right) x_0 y^2_0- \nonumber \\&& 3 \left(5 \mu ^4+33 \mu
^2+30\right) y^3_0 \Big\}+ {\cal O}\left(\frac{QL}{|X|^2}\right)+{\cal
O}\left(\gamma^2 L^3 Q |X|^2\right) \Bigg),
\end{eqnarray}
where $x_0$ and $y_0$ are the functions of $X$ and $Y$ defined in
(\ref{kappaboundary}). Since we consider here only the result (\ref{PYXapprox}) for
the large $\mathrm{SNR}$ limit we imply that $|X|^2 \gg QL$. Note that
 the conditional  PDF $P[Y|X]$ was already derived in \cite{tdyt03} in the form of an infinite series. Our result (\ref{PYXapprox}) for the function $P[Y|X]$
is the analytic summation of this series in the limit of large $\mathrm{SNR}$ and
intermediate power region
\begin{eqnarray}\label{region}
QL \ll P \ll \left( Q L^3 \gamma^2 \right)^{-1}.
\end{eqnarray}
The accuracy ${\cal O}\left(\gamma^2 L^3 Q |X|^2\right)$ in Eq.~(\ref{PYXapprox})
appears in the calculation of the path-integral as a result of neglecting higher
powers of the field $\tilde{\psi}$, see  Eq.~(\ref{deltaS1}) in
Appendix~\ref{Appendix2}. One can show that the normalization condition  $\int DY
P[Y|X]=1$ is fulfilled. Also one can check that the distribution (\ref{PYXapprox})
obeys the following important property
\begin{eqnarray}\label{PYXlimitQzero}
&& \lim_{Q \rightarrow 0} P[Y|X]=\delta\Big(Y-\Psi_{0}(L)\Big).
\end{eqnarray}
The expression (\ref{PYXlimitQzero}) is nothing else, but  the deterministic limit
of $P[Y|X]$ in the absence of noise.  Also Eq.~(\ref{PYXapprox}) has the correct
limit  for the linear channel ($\gamma \to 0$):
\begin{eqnarray}\label{PYXlinear}
P^{(0)}[Y|X]=\frac{e^{- |Y-X|^2/{Q L}}} {\pi Q L }\,,
\end{eqnarray}
that is nothing else but the conditional PDF for the linear nondispersive channel with AWGN.
%=================
\subsection {Output signal PDF $P_{out}[Y]$}

Now we proceed to calculate the probability density function of the output signal $P_{out}[Y]$. Let us consider the integral, see Eq. (\ref{Pout}),
\begin{eqnarray}\label{si}
P_{out}[Y]=\int {\cal D} X P[Y|X] P_{X}[X],
\end{eqnarray}
where the function $P_{X}[X]$ is a smooth function with a scale of variation $P$
which is much greater than $QL$. In that case we can calculate the integral
(\ref{si}) with accuracies ${\cal O}\left({1}/{{\mathrm{SNR}}}\right)$ and ${\cal
O}\left(\gamma^2 L^3 Q P\right)$ by Laplace's method \cite{Lavrentiev:1987}, see
Appendix \ref{Appendix3}. The result has the form:
\begin{eqnarray}\label{limitsmallQ}
P_{out}[Y]=\int {\cal D} X P[Y|X]  P_{X}[X]=  P_{X}\left[Ye^{-i\gamma
|Y|^2L}\right]\left(1+ {\cal O}\left({1}/{{\mathrm{SNR}}}\right)+{\cal O}\left(\gamma^2 L^3
Q P\right)\right).
\end{eqnarray}
The main term of this result (\ref{limitsmallQ}) can be easily obtained from the
following reasoning. The function $P[Y|X]$, see Eq.~(\ref{PYXapprox}), varies on a
scale of order $QL$ which is much less than the scale of $P_{X}[X]$ (the function
$P[Y|X]$ is essentially narrower than the function $P_{X}[X]$). Also $P[Y|X]$ has
the delta-function limit (\ref{PYXlimitQzero}) and therefore in the integral
(\ref{si}) it can be replaced with the delta-function. Note that to obtain the
result (\ref{limitsmallQ}) we do not require the limit $Q \rightarrow 0$ but only
the relation between the scales ${P}$ and $QL$ to be satisfied. In what follows we
will omit the terms ${\cal O}(\ldots)$ for brevity and restore them in the final
results. For the case of the distribution $P_X[X]$ which depends only on $|X|$ we
have $P_{out}[Y] =P_X[|Y|]$. For such distributions we can calculate corrections to
(\ref{limitsmallQ}) in the parameter $QL$ in any order in $QL$.

Let us restrict our consideration in the remainder of this Section to the
distributions $P_{X}[X]$ depending only on $|X|$. We can use the $P[Y|X]$ found in
Ref.~\cite{tdyt03}, see Eqs.~(11)--(13) therein. In this case $P_{out}[Y]$ is a
function of $|Y|=\rho^\prime$
\begin{eqnarray}\label{tur}
{P}_{out}[\rho']=  \dfrac{2 e^{-\dfrac{\rho'^2}{QL}}}{QL}\int^{\infty}_{0}d\rho \rho
e^{-\dfrac{\rho^2}{QL}} I_{0}\left( \frac{2\rho \rho'}{QL}\right){P}_{X}[\rho],
\end{eqnarray}
where $I_0(z)$ is the modified Bessel function of the first kind. Using this
representation we can obtain the simple relation for ${P}_{out}[\rho']$ calculation
in the perturbation theory in $QL$. To this end we perform the zero order Hankel
transformation \cite{Lavrentiev:1987}:
\begin{eqnarray}\label{Hankeltransform}
\hat{P}[k]=\int^{\infty}_{0}d\rho \rho J_{0}(k \rho) P_{X}[\rho].
\end{eqnarray}
of both sides of  Eq.~(\ref{tur}), then we use the standard integral
\cite{GradshteinRyzik} with Bessel and modified Bessel functions
\begin{eqnarray}
&& \int^{\infty}_{0}dz z e^{-p z^2} J_{\nu}\left(b z\right) I_{\nu} (c z)=\frac{1}{2
p} J_{\nu}\left(\frac{b c}{2 p}\right) e^{\frac{c^2-b^2}{4 p}}, \nonumber
\end{eqnarray}
and arrive at the simple relation between the Hankel images
\begin{eqnarray}\label{Hankel1}
\hat{P}_{out}[k]=e^{-k^2 \frac{QL}{4}} \hat{P}[k].
\end{eqnarray}
Performing the inverse Hankel transformation
\begin{eqnarray}\label{inverseHankeltransform}
{P}_{X}[\rho]=\int^{\infty}_{0}dk k J_{0}(k \rho) \hat{P}[k],
\end{eqnarray}
we obtain
\begin{eqnarray}\label{Hankel2}
{P}_{out}[\rho]=e^{\frac{QL}{4} \Delta_{\rho}}  {P}_{X}[\rho],
\end{eqnarray}
where $\Delta_{\rho}=\frac{d^2}{d \rho^2}+\frac{1}{\rho}\frac{d}{d\rho}$ is the
two-dimensional radial Laplace operator.  From the relation (\ref{Hankel2}) the
problem of finding $(QL)^{n}$ corrections to ${P}_{out}[\rho]$ reduces to the
exponent expansion and straightforward calculations of the action of the
differential operator $\Delta_{\rho}^n$ on ${P}_{X}[\rho]$.

Let us consider the widely used example of the modified Gaussian distribution
\begin{eqnarray} \label{PXbeta}
P_X^{(\beta)}[\rho]= \frac{ \exp\left\{- {\beta \rho^2}/{(2 P)}\right\}
\rho^{\beta-2}}{\pi \Gamma\left({\beta}/{2}\right) \left(2P/\beta\right)^{\beta/2}}.
\end{eqnarray}
For $\beta>0$ the distribution $P_X^{(\beta)}[\rho]$ is normalized to unity, $2\pi
\int^{\infty}_{0} d \rho \rho P_X^{(\beta)}[\rho] = 1$, and has the average power
$P$, $2\pi \int^{\infty}_{0} d \rho \rho^3 P_X^{(\beta)}[\rho] = P$. The
distribution $P_X^{(\beta)}[X]$ generalizes the half-Gaussian distribution
(\ref{halfgauss}) for $\beta=1$ and the Gaussian for $\beta=2$:
\begin{eqnarray}\label{PX}
&&P_X^{(2)}[X]=\frac{1}{\pi P} e^{- |X|^2/P}.
\end{eqnarray}
Inserting (\ref{PX}) into Eq.~(\ref{tur}) we obtain a standard integral which can be
found in \cite{GradshteinRyzik}. The result for the output signal PDF has the form:
\begin{eqnarray} \label{PoutYbeta}
&& \!\!\!P_{ out}^{(\beta)}[Y]= _1\!\!F_1 \left(\frac{\beta}{2};1; \frac{|Y|^2
2P}{QL(2P+\beta QL)}\right)  \times \nonumber \\&&  \frac{\exp\{-|Y|^2/QL\}}{\pi Q L
}\left( \frac{\beta QL}{2 P+\beta QL}\right)^{\beta/2},
\end{eqnarray}
where $_1F_1(\frac{\beta}{2};1;z)$ is the confluent hypergeometric function that
reduces to $e^z$ for the Gaussian case and to $e^{z/2} I_{0}(z/2)$ for the
half-Gaussian case:
\begin{eqnarray} \label{Poutgauss}
&& P_{out}^{(2)}[Y]=\frac{1}{\pi(P+Q L)}\exp\left\{- \frac{|Y|^2}{P+Q L}\right\},
\end{eqnarray}

\begin{eqnarray} \label{Pouthalfgauss}
&& P_{out}^{(1)}[Y]= \frac{\exp\{-|Y|^2/(2P+QL)\}}{\pi |Y| \sqrt{\pi(2P+QL)}} \times
  \nonumber \\&& \sqrt{\frac{\pi |Y|^2}{QL}} e^{- \frac{|Y|^2
P}{QL(2P+QL)}}I_0\left(\!\!\frac{|Y|^2 P} {QL(2P+QL)}\right).
\end{eqnarray}
Note that the result for $P_{out}^{(1)}[Y]$ in Ref.~\cite{Mansoor:2011}, see
Eq.~(38) therein, for the half-Gaussian distribution is incorrect. To demonstrate
the general result of Eq. (\ref{limitsmallQ}) let us consider
Eq.~(\ref{Pouthalfgauss}) in the case $QL \ll |Y|^2 \sim P$. For the case one can obtain:
\begin{eqnarray} \label{Pouthalfgauss-1}
&& P_{out}^{(1)}[Y]= P_{X}^{(1)}[|Y|]\left(1+{\cal
O}\left(\frac{1}{\mathrm{SNR}}\right)\right).
\end{eqnarray}
The result (\ref{Pouthalfgauss-1}) coincides with Eq.~ (\ref{limitsmallQ}) with the
accuracy ${\cal O}\left(\gamma^2 L^3 Q P\right)$.

%==========================================Optimal input distribution
\section{Optimal input signal distribution at large $\mathrm{SNR}$}
\label{section3}

The optimal input signal distribution at large $\mathrm{SNR}$ can be found
calculating the mutual information (\ref{MI}) and then maximizing the result with respect to the input
signal distribution function $P_{X}[X]$. Let us start from the calculation of the output
signal entropy $H[Y]$, see Eq. (\ref{entropies}), at large $\mathrm{SNR}$. When the
parameter $\mathrm{SNR} \gg 1$ we can substitute $P_{X}\left[Y \exp\left\{- i \gamma
|Y|^2 L\right\}\right]$ instead of $P_{out}[Y]$ due to the relation
(\ref{limitsmallQ}):
\begin{eqnarray}\label{HY}
H[Y]&=&- \int^{2\pi}_0 d \phi \!\! \int^{\infty}_{0}d \rho' \rho' P_{X}
\left[\rho' e^{i\phi}\right] \times\nonumber\\
&&\log P_{X}\left[\rho' e^{i\phi}\right] +{\cal
O}\left(\frac{1}{\mathrm{SNR}}\right)+{\cal O}\left(\gamma^2 L^3 Q P\right).
\end{eqnarray}
In order to obtain Eq.~(\ref{HY}) we have performed the change of the integration variable
${\phi}=\phi^{(Y)}+ \gamma |Y|^2 L$. One can see that the output signal entropy
(\ref{HY}) coincides with the input signal entropy in the leading order in
$1/\mathrm{SNR}$ and  $\gamma^2 L^3 Q P$.

The conditional entropy $H[Y|X]$ can be calculated by substitution of $P[Y|X]$ in
the form of Eq.~(\ref{PYXapprox}) into Eq.~(\ref{condentropy}). After the
substitution we change the integration variables ${\cal D }Y \equiv d Re Y d Im Y$
to $d x_0 d y_0$. Then we perform integration over $x_0$, $y_0$ and obtain
\begin{eqnarray}\label{HYX}
&&\!\!\!\!\!\!\!\!\!\!\!\!H[Y|X]=1+\log({\pi QL})+ \nonumber \\
&&\!\!\!\!\!\!\!\!\!\!\!\! \frac{1}{2}\int^{2\pi}_0 d \phi^{(X)} \int^{\infty}_{0}d
\rho \rho P_{X}\left[\rho e^{i \phi^{(X)}}\right] \log\left(1+\frac{\gamma^2 L^2}{3}
\rho^4 \right) +{\cal O}\left(\frac{1}{\mathrm{SNR}}\right)+{\cal O}\left(\gamma^2
L^3 Q P\right),
\end{eqnarray}
where the first two terms come from the Gaussian type integrals over ${\cal D} Y$ in
the conditional entropy definition (\ref{condentropy}) and the normalization factor
$\pi Q L$ in Eq. (\ref{PYXapprox}). The third term in Eq. (\ref{HYX}) comes from the
normalization factor $ \sqrt{1+\mu^2/3}$, see Eq. (\ref{PYXapprox}). Note that there
are no terms which are ${\cal O}\left({1}/\sqrt{\mathrm{SNR}}\right)$ in Eqs.~
(\ref{HY}) and  (\ref{HYX}). Indeed, the integrals with the odd powers of $x_0$ and
$y_0$ vanish when integrating over $x_0$, $y_0$ in Eq.~ (\ref{condentropy}) for
$H[Y|X]$.

To find the optimal distribution $P_X^{opt}[X]$ normalized to unity and having a fixed
average power $P$ one should solve the variational problem for the functional
$J[P_{X},\lambda_{1},\lambda_{2}]$
\begin{eqnarray}\label{functional}
&& J[P_{X},\lambda_{1},\lambda_{2}]=H[Y]-H[Y|X]-\lambda_1\left( \int {\cal D}X
P_{X}[X] -1\right)- \nonumber \\&& \lambda_2\left( \int {\cal D}X P_{X}[X] |X|^2 -P
\right),
\end{eqnarray}
where $\lambda_{1,2}$ are Lagrange multipliers. We substitute $H[Y]$ and $H[Y|X]$
from Eqs.~(\ref{HY}) and (\ref{HYX}) to (\ref{functional}),  perform the variation
of the  functional $J[P_{X},\lambda_{1},\lambda_{2}]$ over $P_{X}[X]$, $\lambda_{1}$,
$\lambda_{2}$, and write the Euler-Lagrange equations $\delta
J[P_{X},\lambda_{1},\lambda_{2}]=0$:
\begin{eqnarray}\label{Euler-Lagrange-1}
&&  \int {\cal D}X P_{X}[X] =1,  \\&& \label{Euler-Lagrange-2} \int {\cal D}X
P_{X}[X] |X|^2 =P,
\\&& \label{Euler-Lagrange-3} -1-\log P_{X}[X]-\frac{1}{2}\log\left(1+\frac{\gamma^2 L^2}{3} |X|^4
\right)-\lambda_{1} -\lambda_{2}|X|^2=0.
\end{eqnarray}
The solution $P_X^{opt}[X]$ of Eqs.~(\ref{Euler-Lagrange-1})-(\ref{Euler-Lagrange-3})
referred to as the ``optimal'' distribution depends only on $|X|$ and has the form:
\begin{eqnarray}\label{optimalPDF}
&&P_X^{opt}[X]=N_{0}(P)\frac{\exp\left\{- \lambda_{0}(P)
|X|^2\right\}}{\sqrt{1+\gamma^2 L^2 |X|^4/3}},
\end{eqnarray}
where functions $N_{0}(P)$ and  $\lambda_{0}(P)$ are determined from the conditions
~(\ref{Euler-Lagrange-1}), (\ref{Euler-Lagrange-2}):
\begin{eqnarray}\label{normalization1}
\int {\cal D} X P_X^{opt}[X]= 2\pi N_{0}(P)\int^{\infty}_{0} \frac{d\rho \,  \rho \,
e^{- \lambda_{0}(P) \rho^2}}{\sqrt{1+\gamma^2 L^2 \rho^4/3}} =1,
\end{eqnarray}
\begin{eqnarray}\label{normalization2}
\!\!\int \!\! {\cal D} X P_X^{opt}[X] |X|^2 \!\! = 2\pi N_{0}(P) \!\!
\int^{\infty}_{0} \!\! \frac{d\rho \, \rho^3 \! e^{- \lambda_{0}(P)
\rho^2}}{\sqrt{1+\gamma^2 L^2 \rho^4/3}} =P.
\end{eqnarray}
In a parametric form this dependance reads
\begin{eqnarray}\label{Nandlambda}
&& \lambda_{0}(P)= \frac{\gamma L}{ \sqrt{3}\, }\, \alpha ,\qquad
N_{0}(P)=\frac{\gamma L}{\pi \sqrt{3}\, G(\alpha)},
\end{eqnarray}
here $G(\alpha)=\int^{\infty}_{0} \frac{dz \,\, e^{-\alpha z}}{\sqrt{1+ z^2}}=
\frac{\pi}{2}\Big\{ H_{0}(\alpha)-Y_{0}(\alpha)\Big\}$ with $Y_{0}(\alpha)$ and
$H_{0}(\alpha)$  being the  Neumann and Struve functions of zero order,
respectively. The parameter $\alpha(P)>0$ emerges as the real solution of the
nonlinear equation $ \frac{d}{d \alpha} \log G(\alpha)= - \gamma LP/\sqrt{3} $,
{which comes from Eqs. (\ref{normalization1}) and (\ref{normalization2}).} Let us
emphasize that the optimal distribution obtained here, $P_X^{opt}[X]$
(\ref{optimalPDF}), is different from the half-Gaussian distribution, see
Eq.~(\ref{PXbeta}) for $\beta=1$, whereas in the Ref.~ \cite{Mansoor:2011} the
half-Gaussian distribution was considered as optimal. For sufficiently large values
of the power $P$, such that $\log(\gamma PL)\gg1$,  we can simplify
(\ref{Nandlambda}) using the asymptotic expansions of $Y_{0}(\alpha)$ and
$H_{0}(\alpha)$ at small $\alpha$, see \cite{GradshteinRyzik}:
\begin{eqnarray}\label{Nandlambdaas}
 \lambda_{0}(P)&\approx& \frac{1-\log \log( C \tilde{\gamma})/\log(C \tilde{\gamma})}{P
\log(C \tilde{\gamma})}, \nonumber
\\ N_{0}(P) &\approx& \frac{\tilde{\gamma}}{\pi}\lambda_{0}(P),
\end{eqnarray}
where $C= 2  e^{-\gamma_{E}}$ and $\tilde{\gamma}=\gamma L P/\sqrt{3}$. At small
$P$, the parameter $\tilde{\gamma}\ll1$, the solution of the  Eqs.
(\ref{normalization1}) and (\ref{normalization2}) has the form:
\begin{eqnarray}\label{NandlambdaasSmallGamma}
\lambda_0(P)=\frac{1}{P}\left(1-2\tilde{\gamma}^2\right),\,\, N_0(P)&=&\frac{1}{\pi
P}\left(1-\tilde{\gamma}^2\right).
\end{eqnarray}
It is worth noting that at $\tilde{\gamma} \to 0$ our distribution
(\ref{optimalPDF}) approaches the Gaussian distribution (\ref{PX}) that is known to
be optimal for the linear channel \cite{Shannon:1948}.

%==========================================The mutual information
\section{The mutual information}
\label{section4}

In this Section we present the entropies and the mutual information for
$P_X^{(\beta)}[X]$  and $P_X^{opt}[X]$ and compare our new results with those already
known.

\begin{figure}[h]
\begin{center}
\includegraphics[width=6.7cm]{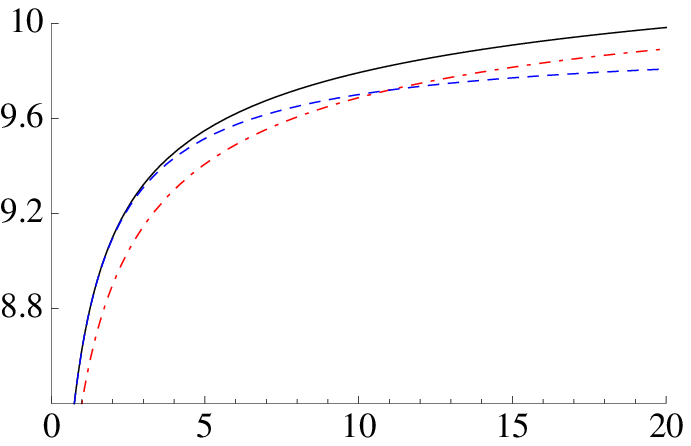} \hspace*{1cm}
\includegraphics[width=7cm]{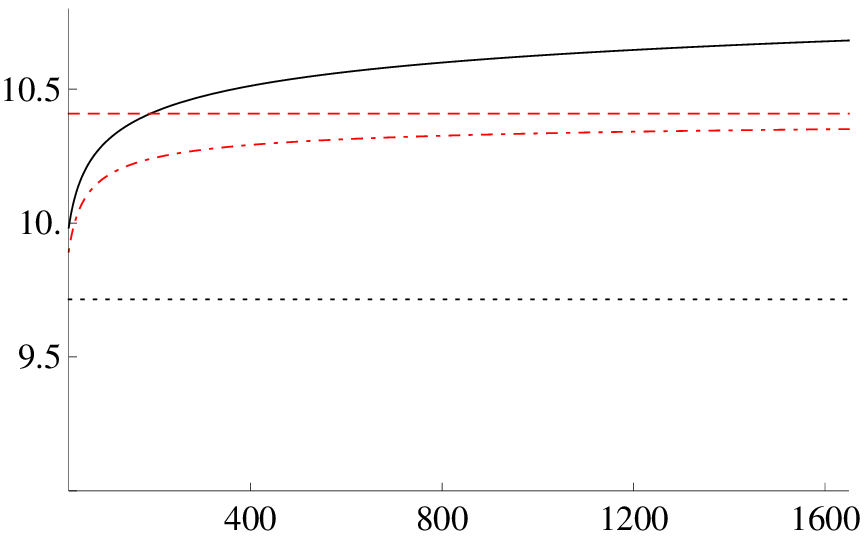}
\begin{picture}(0,0)
\put(-96,-10){\text{$P$[mW]}}
\put(-210,29){\rotatebox{90}{\text{$I_{P[X]}$[nat/symb.]}}}
\put(-175,120){\text{(b)}}

\put(-350,-10){\text{$P$[mW]}}
\put(-444,29){\rotatebox{90}{\text{$I_{P[X]}$[nat/symb.]}}}
\put(-407,120){\text{(a)}}
\end{picture}
\end{center}
\caption{\label{figure1} The mutual information for various input PDFs  as a
function of input average power $P$ for the parameters $Q=1.5 \times 10^{-7} \,
\mathrm{mW}\, \mathrm{km}^{-1} $, $\gamma = 10^{-3} \, \mathrm{mW}^{-1}
\mathrm{km}^{-1}$, $L=1000 \,\mathrm{km}$. (a): The solid black line,  blue dashed
line, red dashed dotted line correspond to the optimal PDF $P_X^{opt}[X]$, Gaussian
PDF $P_X^{(2)}[X]$, and half-Gaussian PDF $P_X^{(1)}[X]$, respectively. (b): The
solid black line corresponds to $I_{P_X^{opt}[X]}$, see Eq.~(\ref{optimalMI}); the red
dashed dotted line corresponds to the mutual information for the half-Gaussian
distribution $I_{P_X^{(1)}[X]}$, see Eq.~(\ref{MutualGauss}) for $\beta=1$; the red
dashed horizontal line corresponds to our limit (\ref{MIhgasymp}) at
$\tilde{\gamma}\gg 1$ for the half-Gaussian distribution; the black dotted
horizontal line corresponds to the result \cite{Mansoor:2011}, see
Eq.~(\ref{capacityyousefi}).}
\end{figure}

Substituting the expression (\ref{optimalPDF}) for $P_X^{opt}[X]$ in equations
(\ref{HY})-(\ref{HYX}) and using the definition (\ref{MI}) we obtain   the mutual
information for the optimal distribution in the leading order  in $1/\mathrm{SNR}$:
\begin{eqnarray}\label{optimalMI}
\!\!\!I_{P_X^{opt}[X]}= P \lambda_{0}(P) - \log N_{0}(P) -\log(\pi e Q L) +{\cal
O}\left(\frac{1}{\mathrm{SNR}}\right)+{\cal O}\left(\gamma^2 L^3 Q P\right),
\end{eqnarray}
which gives a capacity estimate in a wide range (\ref{region}) of the average power
$P$. The mutual information (\ref{optimalMI}) is depicted by the black solid line in
Fig.~\ref{figure1} as a function of power $P$ for the following parameters: $Q=1.5
\times 10^{-7} \, \mathrm{mW}\, \mathrm{km}^{-1} $, $\gamma = 10^{-3} \,
\mathrm{mW}^{-1} \mathrm{km}^{-1}$, $L=1000 \,\mathrm{km}$. For these realistic
parameters, the power range (\ref{region}) is actually very wide:
\begin{eqnarray}\label{regionnum}
1.5 \times 10^{-4} \mathrm{mW} \ll P \ll 0.66  \times 10^{4} \mathrm{mW}.
\end{eqnarray}

%========
There is no simple analytical form for $N_{0}(P)$ and $\lambda_{0}(P)$ therefore to
plot Fig.~\ref{figure1} and Fig~\ref{figure2} (see below) we calculated
$\lambda_{0}(P)$ and $ N_{0}(P) $ numerically. For large and small values of the
parameter $\tilde{\gamma}$ we can use the solutions in Eqs. (\ref{Nandlambdaas}) and
(\ref{NandlambdaasSmallGamma}), respectively. At small $\tilde{\gamma}= \gamma L P
/\sqrt{3}$ we obtain
\begin{eqnarray}\label{optimalMIsmall}
\!\!\!I_{P_X^{opt}[X]}= \log\left(1+\mathrm{SNR}\right)-\tilde{\gamma}^2 +  {\cal
O}(\tilde{\gamma}^3) + {\cal O}\left(\frac{1}{{\mathrm{SNR}}}\right),
\end{eqnarray}
which is simply the Shannon capacity, $\log\left(1+\mathrm{SNR}\right)$, of the
linear AWGN channel (\ref{CapacityShannon}) with the first nonlinear correction. In
Eq.~(\ref{optimalMIsmall}) the unity in the logarithm is beyond the accuracy of our calculation but
we keep it to bring to notice that the derived expressions (\ref{HYX}) and
(\ref{optimalMIsmall}) have the correct limit when the parameter $\gamma$ tends to
zero (in contrast to the Eq.~(35) in Ref.~\cite{Mansoor:2011}). In
Eq.~(\ref{optimalMIsmall}) we omit the accuracy ${\cal O}\left(\gamma^2 L^3 Q
P\right)$  since the parameter $\gamma^2 L^3 Q P$ is of order of
$\tilde{\gamma}^2/\mathrm{SNR}$. In the second power sub-interval $(\gamma L)^{-1}
\ll P \ll \left( Q L^3 \gamma^2 \right)^{-1}$, using Eq.~(\ref{Nandlambdaas}) one
can see that the mutual information increases very slowly (loglog) with $P$
\begin{eqnarray}\label{optimalMIlarge}
&&I_{P_X^{opt}[X]}= -\log\left({QL^2\gamma}\right)-1+\frac{\log
3}{2}+\log\log\left(\frac{C \gamma L P}{\sqrt{3}}\right)+ \nonumber
\\&& \frac{1}{\log\left({C
\gamma L P}/{\sqrt{3}}\right)}\left[ \log\log\left(\frac{C \gamma L
P}{\sqrt{3}}\right)+1 - \frac{\log\log\left({C \gamma L
P}/{\sqrt{3}}\right)}{\log\left({C \gamma L P}/{\sqrt{3}}\right)} \right] +\nonumber
\\&& {\cal O}\left( 1/\log^2(\gamma L P)\right)+{\cal
O}\left(1/\mathrm{SNR}\right)+{\cal O}\left(\gamma^2 L^3 Q P\right),
\end{eqnarray}
as opposed to the constant behavior of the mutual information for Gaussian-like
distributions of an input signal, see formulae (\ref{MIbetaasymp}) and
(\ref{MIhgasymp}) below.

In the remainder of this Section we perform an analysis of the mutual information
for the  distribution $P_X^{(\beta)}[X]$, see Eq.~(\ref{PXbeta}), generalizing the
half-Gaussian distribution (\ref{halfgauss}) (see, for example
Ref.~\cite{Mansoor:2011}) and the Gaussian input PDF (\ref{PX}). In the leading
order in $1/\mathrm{SNR}$ from  (\ref{HY}) we obtain
\begin{eqnarray}\label{HYbeta}
&&\!\!\!\!\!\!\!\!\!\! H_{\beta}[Y]= \log\left(  P \frac{2\pi}{\beta}
\Gamma\left(\frac{\beta}{2} \right)\right)+ \frac{\beta}{2}+
\frac{2-\beta}{2}\psi\Big(\frac{\beta}{2}\Big),
\end{eqnarray}
where $\psi(z)$ is the digamma function $\psi(z)=\Gamma'(z)/\Gamma(z)$, where
$\psi(1)=-\gamma_{E}$ and $\psi(1/2)=-\gamma_{E}-2 \log(2)$. The substitution of Eq.
(\ref{PXbeta}) into Eq.~(\ref{HYX}) gives
\begin{eqnarray}\label{HYXbeta}
&&\!\!\!\!\!\!H_{\beta}[Y|X]= \log\Big(\pi e Q L\Big)+ \nonumber \\&&
\frac{1}{2\Gamma\left(\frac{\beta}{2} \right)} \int^{\infty}_{0} d\tau
e^{-\tau}\tau^{\beta/2-1} \log\left(1+\frac{4\tilde{\gamma}^2}{\beta^2}\tau^2\right)
\end{eqnarray}
with at least ${{\cal O}(1/{\mathrm{SNR}})}$ accuracy. The integral in Eq.
(\ref{HYXbeta}) can be calculated analytically using Ref. \cite{GradshteinRyzik},
however, the result of the integration is cumbersome, hence we do not present it
here. One can easily obtain the mutual information $I_{P_X^{(\beta)}[X]}$ by
subtracting Eq.~(\ref{HYXbeta}) from Eq. (\ref{HYbeta}):
\begin{eqnarray}\label{MutualGauss}
I_{P_X^{(\beta)}}[X]&=&\log\mathrm{SNR}+\log\left(\frac{2\Gamma\left(\beta/2 \right)}{\beta}  \right)-\nonumber \\
&& \frac{1}{2\Gamma\left(\frac{\beta}{2} \right)} \int^{\infty}_{0} d\tau
e^{-\tau}\tau^{\beta/2-1}
\log\left(1+\frac{4\tilde{\gamma}^2}{\beta^2}\tau^2\right)+\nonumber
\\
&&  \frac{\beta-2}{2}\left(1-\psi\left(\frac{\beta}{2}\right)\right)+ {{\cal
O}\left(\frac{1}{{\mathrm{SNR}}}\right)+{\cal O}\left(\gamma^2 L^3 Q P\right)}.
\end{eqnarray}
The mutual information is depicted in Fig.~\ref{figure1}(a) for the Gaussian
distribution by the blue dashed line, and for the half-Gaussian by the red dashed
dotted line. One can see that at small $P$ the mutual information for the Gaussian
distribution is greater than that of the half-Gaussian, whereas at $P>11
\mathrm{mW}$ the mutual information is greater for the half-Gaussian distribution.
Note that $I_{P_X^{opt}[X]}$ is greater than $I_{P_X^{(\beta)}[X]}$ for all values of
$P$,  as it should be. At  $\tilde{\gamma}\gg 1$ the mutual information
$I_{P_X^{(\beta)}[X]}$ takes the form
\begin{eqnarray}\label{MIbetaasymp}
I_{P_X^{(\beta)}[X]}\!\!&=&\!\! -\log\left(Q L^2 \gamma\right)-
\frac{2-\beta}{2}+\frac{\log 3}{2}-\frac{\beta}{2}\psi \left( \frac{\beta}{2}
\right) +
\nonumber \\
&& \log \left(\Gamma \left( \beta/2 \right)\right)+ {{\cal
O}\left(\frac{1}{{\mathrm{SNR}}}\right)+{\cal O}\left(\gamma^2 L^3 Q P\right)}.
\end{eqnarray}
One can see that at large $\mathrm{SNR}$  $I_{P_X^{(\beta)}[X]}$ goes to a constant
in the interval of power $P$ considered, and this constant depends on the noise
power $QL$. We remind that $I_{P_X^{opt}[X]}$ increases as $\log\log P$ in the region
under consideration. The mutual information for the half-Gaussian distribution
(\ref{halfgauss}) in the regime $\tilde\gamma\gg 1$ can be obtained as a particular
case of (\ref{MIbetaasymp}) for $\beta=1$:
%\begin{eqnarray}\label{MIhgasymp}
%&&\!\!\!\!\!\!I_{P_{1}[X]}= \log(\frac{1}{Q L^2 \gamma} )- \frac{1}{2}+\frac{\log 3}{2}+ \nonumber \\&&  \frac{\log \pi}{2} +\frac{\gamma_E}{2}+\log 2 + {\cal O} \left(\frac{1}{\sqrt{\mathrm{SNR}}}\right),
%\end{eqnarray}
\begin{eqnarray}\label{MIhgasymp}
I_{P_X^{(1)}[X]}&=& -\log\left({Q L^2 \gamma}\right)+\log 2+\nonumber\\
&&\frac{\log 3\pi-1+\gamma_E}{2}
 +{ {\cal O}\left(\frac{1}{{\mathrm{SNR}}}\right)+{\cal
O}\left(\gamma^2 L^3 Q P\right)}.
\end{eqnarray}
Comparing our expression (\ref{MIhgasymp}) with the result (40) of
Ref.~\cite{Mansoor:2011} we have an extra term $+\log 2$  due to our more accurate
calculation of $H[Y|X]$. Our result (\ref{MIhgasymp}) and the result of
Ref.~\cite{Mansoor:2011}, see Eq.~(\ref{capacityyousefi}), are presented in
Fig.~\ref{figure1}(b). In Fig.~\ref{figure1}(b) one can see that the mutual
information (\ref{optimalMI}) for the optimal distribution exceeds the limit
(\ref{MIhgasymp}) at $P \sim 190$mW. At this power the difference between the limit
(\ref{MIhgasymp}) and  $I_{P_X^{(1)}[X]}$ evaluated on the base of
Eq.~(\ref{MutualGauss}) with $\beta=1$ is of order of $1.5\%$ and getting smaller at
higher $P$.

Since we have now found $P_X^{opt}[X]$ in the power region (\ref{region}), we can
calculate an approximation for the capacity of the considered per-sample nonlinear
channel. By definition it coincides with the mutual information expression
(\ref{optimalMI}):
\begin{eqnarray}\label{CapacityOur}
C=I_{P_X^{opt}[X]}.
\end{eqnarray}
Let us emphasize that this result for the capacity has accuracy ${\cal
O}\left(1/{\mathrm{SNR}}\right)+{\cal O}\left(\gamma^2 L^3 Q P\right)$, see
Eq.~(\ref{optimalMI}). The comparison of  the approximation (\ref{CapacityOur}) with
the Shannon capacity of the linear AWGN channel is presented in Fig.~\ref{figure2}.
%=================
\begin{figure}[h]
\begin{center}
\includegraphics[width=7cm]{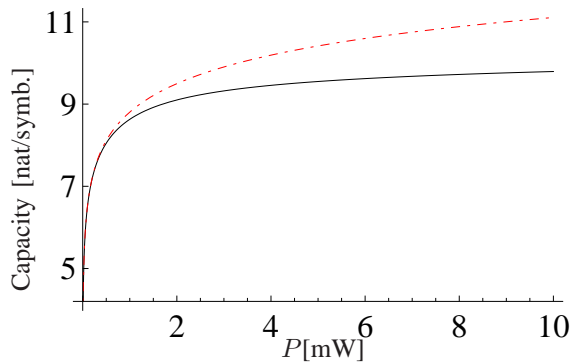}
\begin{picture}(0,0)
\put(-113,-5){\text{$P$[mW]}} \put(-215,20){\rotatebox{90}{\text{Capacity
[nat/symb.]}}}
\end{picture}
\end{center}
\caption{\label{figure2} Shannon capacity and the capacity of the nonlinear channel
$I_{P_X^{opt}[X]}$ for the parameters $Q=1.5 \times 10^{-7} \, \mathrm{mW}\,
\mathrm{km}^{-1} $, $\gamma = 10^{-3} \, \mathrm{mW}^{-1} \mathrm{km}^{-1}$, $L=1000
\,\mathrm{km}$. The black solid line corresponds to $I_{P_X^{opt}[X]}$, see
Eq.~(\ref{optimalMI}), the red dashed-dotted line corresponds to the Shannon limit
$\log[1+\mathrm{SNR}]$. }
\end{figure}
One can see that the Shannon capacity of the linear AWGN channel is always greater
than the approximation (\ref{CapacityOur}) for the nondispersive nonlinear fiber
channel for the considered region of $P$.

%==========================================Conclusion
\section{Conclusion}
\label{section5}

We have developed a new approach to the calculation of the conditional PDF via the
path-integral representation (\ref{QuasiclassProbabInitial}) at large $\mathrm{SNR}$
for the intermediate power region (\ref{region}).  This may be an especially useful
technique for complex nonlinear channels in which the calculation of the conditional
PDF is technically challenging. Applying our method to the per-sample nondispersive
nonlinear fiber channel, we derived compact analytical expressions for the
conditional PDF, conditional entropy and the entropy of the output signal for
different input signal PDFs $P_X[X]$. Moreover, we solved the variational problem on
$P_{X}[X]$ maximizing the mutual information in the leading order in
$1/\mathrm{SNR}$ in the power region (\ref{region}). That allows us to find the
optimal input signal distribution (\ref{optimalPDF}) and the approximation for the
channel capacity (\ref{optimalMI}) up to ${1/{\mathrm{SNR}}}$ and $\gamma^2 L^3
Q P$ corrections in the  power interval $QL\ll P\ll (\gamma^2 QL^3)^{-1}$, which is
extremely wide for realistic parameters, see (\ref{regionnum}). The found
distribution $P_X^{opt}[X]$ is different from the half-Gaussian one, and at the zero
nonlinearity $P_X^{opt}[X]$ approaches the Gaussian distribution. We demonstrated that
the approximation found for the capacity of the channel considered here
(\ref{CapacityOur}) is always greater than the mutual information calculated for
Gaussian and half-Gaussian distributions, and lower than the Shannon capacity of the
linear AWGN channel.

%==========================================Acknowledgments

\emph{\it Acknowledgments} The calculations of the next-to-leading order corrections in $1/\mathrm{SNR}$
have been performed with support of the Russian Science Foundation (RSF) (grant No.
16-11-10133). Part of the work (Section III) was supported by the Russian Scientific
Foundation (RFBR), Grant No. 16-31-60031/15. A.~V.~Reznichenko thanks the President
program for support. The work of S.~K.~Turitsyn was supported by the grant of
the Ministry  of Education and Science of the Russian Federation (agreement No.
14.B25.31.0003) and the EPSRC project UNLOC.

%==========================================Appendix
\appendix

\subsection{The classical solution $\Psi_{cl}$ and the action
$S[\Psi_{cl}]$.}\label{Appendix1}

In Ref.~\cite{Terekhov:2014} we have shown that in the case $\mathrm{SNR}={P}/{QL}
\gg 1$  the conditional probability can be written in the form:
\begin{eqnarray}
\label{AQuasiclassProbabInitial} P[Y|X] = \exp \left\{-\frac{S[\Psi_{cl}(z)]}{Q}
\right\}  \!\!\int \limits_{{\psi}(0)=0}^{{\psi}(L)=0}\!\!\!\!\! {\cal D}{\psi}
\,\exp \left\{-\dfrac{S[\Psi_{cl}(z)+{\psi}(z)]-S[\Psi_{cl}(z)]}{Q}\right\}\,,
\end{eqnarray}
where for the nondispersive model  the effective action reads
\begin{eqnarray}\label{AAction}
S[\psi]= \int\limits_{0}^{L}dz \Big|\partial_{z}\psi-i\gamma |\psi|^2 \psi\Big|^2.
\end{eqnarray}
The action (\ref{AAction}) is associated with the l.h.s. of the nonlinear
Shr\"{o}dinger equation
\begin{align}
& \partial_{z}\psi(z)-i\gamma |\psi(z)|^2 \psi(z)=\eta(z),
\label{AShrodingerequation}
\end{align}
where the noise $\eta(z)$ has the  Gaussian nature:
\begin{eqnarray} \langle \eta (z) \rangle_{\eta} =
0\,, \qquad \langle \eta (z)\bar{\eta} (z^\prime)\rangle_{\eta} = Q
\delta(z-z^\prime)\,. \end{eqnarray}

The measure ${\cal D}{\psi}$  in Eq.~(\ref{AQuasiclassProbabInitial}) is defined  as
\be \label{Ameasure} {\cal D} {\psi}=\lim_{\Delta \rightarrow 0} \Big(
\frac{1}{\Delta \pi Q}\Big)^{N }\prod^{N-1}_{i=1}dRe{\psi}_{i}\,dIm{\psi}_{i}, \ee
here ${\psi}_{i}={\psi}(z_i)$ and $\Delta=\frac{L}{N}$ is the grid space.

Now we consider the difference of actions in the exponent of the path-integral in
Eq.~(\ref{AQuasiclassProbabInitial}).
\begin{eqnarray}\label{AdeltaS}
&&{S[\Psi_{cl}(z)+{\psi}(z)]-S[\Psi_{cl}(z)]}=\nonumber \\&& \int^{L}_{0}dz\Big\{
\left|
\partial_{z}{\psi}-i \gamma(2{\psi} |\Psi_{cl}|^2+\bar{\psi}\Psi^2_{cl}) \right|^2 +2\gamma Im
\left[\left(\partial_{z} \bar{\Psi}_{cl}+i \gamma \bar{\Psi}_{cl} |{\Psi}_{cl}|^2
\right) \left(2 \Psi_{cl} |\psi|^2+\bar{\Psi}_{cl}\psi^2\right) \right]+ \nonumber
\\&& \gamma^2 \left|2 \Psi_{cl} |\psi|^2+ \bar{\Psi}_{cl} \psi^2+\psi
|\psi|^2\right|^2 + 2\gamma Im \left[\left(\partial_{z}\bar{\psi}+i
\gamma(2\bar{\psi} |\Psi_{cl}|^2+{\psi}\bar{\Psi}^2_{cl})  \right) \left(2 \Psi_{cl}
|\psi|^2+ \bar{\Psi}_{cl} \psi^2+\psi |\psi|^2 \right)\right] +\nonumber \\&&
2\gamma Im \left[\left(\partial_{z} \bar{\Psi}_{cl}+i \gamma \bar{\Psi}_{cl}
|{\Psi}_{cl}|^2 \right)\psi |\psi|^2 \right]\Big\}.
\end{eqnarray}

%========================================The classical equation of motion
In Eq.~ (\ref{AQuasiclassProbabInitial}) the function $\Psi_{cl}(z)$ is the solution
of the equation $\delta S[\Psi_{cl}]=0$ (Euler-Lagrange equation) which has the form
\be \label{AclassicalTrajectoryEq} \frac{d^2\Psi_{cl}}{dz^2}-4i\gamma
\left|\Psi_{cl}\right|^2 \frac{d\Psi_{cl}}{dz}-3\gamma^2 \left|\Psi_{cl}\right|^4
\Psi_{cl}=0, \ee with boundary conditions $\Psi_{cl}(0)=X=|X|\exp[{i \phi^{(X)}}]$,
$\Psi_{cl}(L)=Y=|Y|\exp[{i \phi^{(Y)}}]$. It is easy to find the solution of Eq.~
(\ref{AclassicalTrajectoryEq}) in the polar coordinate system:
$\Psi_{cl}(z)=\rho(\zeta)e^{i \theta (\zeta)}$, $\zeta=z/L$. The solution depends on
four real integration constants. We denote them as $E$, $\tilde{\mu}$, $\zeta_0$ and
$\theta_0$. There are two different regimes of the solution: in the trigonometric
regime one has  $E=\frac{k^2}{2}\geq 0$,  and in the hyperbolic regime
$E=-\frac{k^2}{2} \leq 0$. For both cases instead of $E$ we introduce the
non-negative parameter $k=\sqrt{2|E|}$.
\begin{itemize}
  \item In the \textbf{trigonometric case }($E=\frac{k^2}{2}\geq 0$)
   we have the solution for $\tilde{\mu}\geq k \geq 0$:

\be \label{Asolutiontrig}
\begin{split}
&\rho^2(\zeta) =\frac{1}{2 L \gamma}\Big(\tilde{\mu}
+\sqrt{\tilde{\mu}^2-k^2}\cos[2k(\zeta-\zeta_0)] \Big)\,,
\\& \theta (\zeta)= \frac{\tilde{\mu}}{2} (\zeta-\zeta_0) +
\sqrt{\tilde{\mu}^2-k^2}\, \frac{\sin[2k(\zeta-\zeta_0)]}{4 k}+
\arctan\left[(\tilde{\mu}-\sqrt{\tilde{\mu}^2-k^2})\frac{\tan[k(\zeta-\zeta_0)]}{k}\right]+\theta_0.
\end{split}
\ee Here the integration constants $\tilde{\mu}$, $k$ and $\zeta_0$ must be found
from the boundary conditions:
\begin{eqnarray}
\label{Aboundary_1}
&&|X|^2=\rho^2(0)=\frac{1}{2 L \gamma}\Big(\tilde{\mu}
+\sqrt{\tilde{\mu}^2-k^2}\cos[2k\zeta_0] \Big),\\
&& |Y|^2=\rho^2(1)=\frac{1}{2 L \gamma}\Big(\tilde{\mu}
+\sqrt{\tilde{\mu}^2-k^2}\cos[2k(1-\zeta_0)] \Big),\\
&& \phi^{(X)}=\theta (0)= -\frac{\tilde{\mu}}{2} \zeta_0 -\sqrt{\tilde{\mu}^2-k^2}\,
\frac{\sin[2k\zeta_0]}{4 k}-
\arctan\left[(\tilde{\mu}-\sqrt{\tilde{\mu}^2-k^2})\frac{\tan[k\zeta_0]}{k}\right]+\theta_0,\\
&&\phi^{(Y)}=\theta (1)=\frac{\tilde{\mu}}{2} (1-\zeta_0) +
\sqrt{\tilde{\mu}^2-k^2}\, \frac{\sin[2k(1-\zeta_0)]}{4 k}+
\arctan\left[(\tilde{\mu}-\sqrt{\tilde{\mu}^2-k^2})\frac{\tan[k(1-\zeta_0)]}{k}\right]+\theta_0.
\label{Aboundary}
\end{eqnarray}
Then one can find the action \be \label{Aactiontrig} S[\Psi_{cl}(z; E=\frac{k^2}{2},
\tilde{\mu}, \zeta_0, \theta_0)]= \frac{k^2}{2\gamma
L}\left({\tilde{\mu}}-\sqrt{\tilde{\mu}^2-k^2} \,
\frac{\sin[2k(1-\zeta_0)]+\sin[2k\zeta_0]}{2 k} \right). \ee

\item In the \textbf{hyperbolic case} ($E=-\frac{k^2}{2}\leq 0$)  we have the
solution for $k\geq0$ and arbitrary $\tilde{\mu}$ in the following form \be
\label{Asolutionhyp}
\begin{split}
& \rho^2(\zeta)=\frac{1}{2 L \gamma}\Big(-\tilde{\mu}
+\sqrt{\tilde{\mu}^2+k^2}\cosh[2k(\zeta-\zeta_0)] \Big)\,,
\\& \theta (\zeta)= -\frac{\tilde{\mu}}{2} (\zeta-\zeta_0) +
\sqrt{\tilde{\mu}^2+k^2}\, \frac{\sinh[2k(\zeta-\zeta_0)]}{4 k}-
\arctan\left[(\tilde{\mu}+\sqrt{\tilde{\mu}^2+k^2})\frac{\tanh[k(\zeta-\zeta_0)]}{k}\right]+\theta_0,
\end{split}
\ee where $\tilde{\mu}$, $k$, $\zeta_0$, and $\theta_0$ are derived from the same
procedure as in the trigonometric regime. The action reads \be \label{Aactionhyp}
S[\Psi_{cl}(z; E=-\frac{k^2}{2}, \tilde{\mu}, \zeta_0, \theta_0)]=
\frac{k^2}{2\gamma L}\left({\tilde{\mu}}+\sqrt{\tilde{\mu}^2+k^2} \,
\frac{\sinh[2k(1-\zeta_0)]+\sinh[2k\zeta_0]}{2 k} \right). \ee
\end{itemize}
Note, there are two solutions of Eq.~(\ref{AclassicalTrajectoryEq}) with constant
$\rho(z)=\rho(0)\equiv\rho$ obeying only the input boundary condition $\Psi_0(0)=X$.
The first one reads \be \label{Azeronoisesolution} \Psi_0(z)=\rho \exp\left\{i \mu
\frac{z}{L}+ i \phi^{(X)}\right\}, \ee where  $\mu=\gamma L \rho^2 = \gamma L
|X|^2$. This function corresponds to the solution representation
(\ref{Asolutiontrig}) with $k=0$ and $\tilde{\mu}= \mu$ or to the solution
representation (\ref{Asolutionhyp}) with $k=0$ and $\tilde{\mu}= -\mu$. The function
$\Psi_0(z)$ is the solution of the Eq.~ (\ref{AShrodingerequation}) with zero noise
and with the input boundary condition. Furthermore, $\Psi_0(z)$ delivers the
absolute minimum of the action (\ref{AAction}): $S[\Psi_0(z)]=0$. The second
solution of Eq.~(\ref{AclassicalTrajectoryEq}) with constant $\rho(z)$ is the
trigonometric regime (\ref{Asolutiontrig}) case  with $\tilde{\mu}= k= 2 \mu$: \be
\label{Arhoconstsolution} \Psi_{\rho=const}(z)=\rho \exp\left\{ 3i \, \mu
\frac{z}{L}+ i \phi^{(X)}\right\}, \qquad \mu=\gamma L \rho^2 = \gamma L |X|^2. \ee

To find the solution of Eq.~(\ref{AclassicalTrajectoryEq}) one should express the
integration constant through the boundary conditions. Instead, we exploit the fact
that the noise power $QL$ is much less than the input signal power $P=|X|^2\equiv
\rho^2$. In other words, we will find a solution of
Eq.~(\ref{AclassicalTrajectoryEq}) that is close to $\Psi_0(z)$: it is the solution
of Eq.~(\ref{AShrodingerequation}) with zero noise which provides the absolute
minimum of the action $S[\Psi_0(z)]=0$. In that fashion we perform the substitution
in Eq.~(\ref{AclassicalTrajectoryEq}): \be \label{APsiclass}
\Psi_{cl}(z)=(\rho+\varkappa(z))\exp\left\{i \mu \frac{z}{L}+i \phi^{(X)}\right\},
\ee where  the function $\varkappa(z)$ is assumed to be small: $\varkappa(z) \ll
\rho$ for  all configurations of $\Psi_{cl}(z)$ providing $S[\Psi_{cl}(z)]/Q= {\cal
O}(1)$ when $QL$ tends to zero. We have the following equation on $\varkappa(z)$
resulting from the Eq.~(\ref{AclassicalTrajectoryEq}):
\begin{eqnarray}
\label{Akappaequationexact}
&&\frac{d^2\varkappa}{dz^2}-2i\frac{\mu}{L}\frac{d\varkappa}{dz}-4 \frac{\mu^2}{L^2}
Re[\varkappa]= 4i \frac{\mu}{L \rho}\left(\varkappa+\bar{\varkappa}\right)
\frac{d\,\varkappa}{dz}+ \frac{ \mu^2}{L^2
\rho}\left[5\varkappa^2+10|\varkappa|^2+3\bar{\varkappa}^2\right]+\nonumber
\\&& \frac{|\varkappa|^2 \mu}{L^2 \rho^2}\left[ 4 i L
\frac{d\varkappa}{dz}+9{\mu}\bar{\varkappa}+ 14{\mu}{\varkappa}\right]+\frac{3
\mu^2}{L^2 \rho^2}\varkappa^3+ \frac{3 \mu^2}{L^2
\rho^3}|\varkappa|^2\left[3|\varkappa|^2+2\varkappa^2\right]+ \frac{3 \mu^2}{L^2
\rho^4}|\varkappa|^4\varkappa.
\end{eqnarray}

We present $\varkappa(z)$ as a perturbation theory decomposition in powers of
$1/\sqrt{\mathrm{SNR}}$: $\varkappa(z)=\varkappa_1(z)+\varkappa_2(z)+\ldots$, where
$\varkappa_1(z)$ is of $1/\sqrt{\mathrm{SNR}}$ order and provides the leading order
contribution, $\varkappa_2(z)$ is of $1/{\mathrm{SNR}}$ order, and so on.

\begin{itemize}
    \item The linearized equation for the function $\varkappa_1(z)=x_1(z)+i y_1(z)$
    can be obtained from Eq.~(\ref{Akappaequationexact}) by omitting the r.h.s. of
    this equation:
    \be
\label{Akappaequation}
\frac{d^2\varkappa_1}{dz^2}-2i\frac{\mu}{L}\frac{d\varkappa_1}{dz}-4
\frac{\mu^2}{L^2} Re[\varkappa_1]=0. \ee The boundary conditions $\Psi_{cl}(0)=X$
and $\Psi_{cl}(1)=Y\equiv\rho'e^{i\phi^{(Y)}}$ lead to
\begin{eqnarray}\label{Akappaboundary}
&&\varkappa_1(0)=0, \nonumber \\&& \varkappa_1(L)=x_0+iy_0=\rho' e^{i (\phi^{(Y)}-
\phi^{(X)}-\mu)}-\rho.
\end{eqnarray}
The solution $\varkappa_1(z)=x_1(z)+i y_1(z)$ of the linearized boundary problem
(\ref{Akappaequation}), (\ref{Akappaboundary}) is polynomial
\begin{eqnarray}\label{Akappasolution}
&&x_1(z)=\Big(-\mu a_1(X,Y) \frac{z}{L}+ a_2(X,Y) \Big)\frac{z}{L}, \nonumber \\&&
y_1(z)=\Big(-\frac{2}{3}\mu^2 a_1(X,Y) \frac{ z^2}{L^2} + \mu\, a_2(X,Y)
\frac{z}{L}+a_1(X,Y)\Big)\frac{z}{L},
\end{eqnarray}
where coefficients $a_1(X,Y)$ and $a_2(X,Y)$ can be found from the boundary
conditions (\ref{Akappaboundary}) and have the form:
\begin{eqnarray}\label{ACandD}
&&a_1(X,Y)=\frac{-\mu x_0+y_0}{1+\mu^2/3},\qquad  a_2(X,Y)=
\frac{(1-2\mu^2/3)x_0+\mu y_0}{1+\mu^2/3},
\end{eqnarray}
with $x_0=x_0(X,Y)$ and  $y_0=y_0(X,Y)$ being determined from
Eq.~(\ref{Akappaboundary}). In the leading order the action reads
\begin{eqnarray}\label{Aactionkappa}
&&\frac{1}{Q}S\left[\Psi_0(z)+\varkappa_1(z)e^{i \mu \frac{z}{L}+i
\phi^{(X)}}\right]= \frac{1}{Q}\int^{L}_0 dz\left[ \left|\partial_z \varkappa_1- 2 i
\frac{\mu}{L} Re[ \varkappa_1]\right|^2+ {\cal
O}\left(\frac{\varkappa_1^3}{\rho^3}\right) \right] = \nonumber \\&& =
\frac{(1+4\mu^2/3)a^2_1-2\mu a_1 a_2+a^2_2}{QL}+{\cal
O}\left(\frac{1}{\sqrt{\mathrm{SNR}}}\right) = \frac{(1+4\mu^2/3)x^2_0-2\mu x_0
y_0+y^2_0}{QL(1+\mu^2/3)}+{\cal O}\left(\frac{1}{\sqrt{\mathrm{SNR}}}\right).
\end{eqnarray}

    \item Let us proceed to the next-to-leading order correction to $P[Y|X]$. We
should calculate the next approximation $\varkappa_2(z)$ to the solution
(\ref{APsiclass}). Taking into account Eq.~(\ref{Akappaequation}) we present the
equation for $\varkappa_2(z)$ in the form \be \label{Akappaequation2}
\frac{d^2\varkappa_2}{dz^2}-2i\frac{\mu}{L}\frac{d\varkappa_2}{dz}-4
\frac{\mu^2}{L^2} Re[\varkappa_2]= 4i \frac{\mu}{L
\rho}\left(\varkappa_1+\bar{\varkappa}_1\right) \frac{d\,\varkappa_1}{dz}+ \frac{
\mu^2}{L^2 \rho}\left[5\varkappa_1^2+10|\varkappa_1|^2+3\bar{\varkappa_1}^2\right],
\ee where the boundary conditions for $\varkappa_2(z)$ read
$\varkappa_2(0)=\varkappa_2(L)=0$. The solution $\varkappa_2(z)=x_2(z)+ i y_2(z)$ of
Eq.~(\ref{Akappaequation2}) is polynomial  in $z$ and quadratic in $x_0$ and $y_0$:
\begin{equation}\label{Ax2z}% it is correct
\begin{split}
x_2(z)=& -\frac{\mu/\rho }{270 (1+\mu ^2/3)^3  }\Big(1-\frac{z}{L}\Big)
\frac{z}{L}\times
\\& \Big\{\mu \left(2 \mu ^4-15 \mu ^2+585\right) x^2_0+2 \left(13 \mu ^2 \left(\mu
^2+15\right)-180\right) x_0 y_0+ \mu  \left(2 \mu ^2+15\right) \left(5 \mu
^2-9\right) y^2_0 -\\& 5 \left(\mu ^2+3\right)  \frac{z}{L} \left(\mu \left(\mu
^2-15\right) x^2_0-4 \left(\mu ^2-6\right) x_0 y_0+\mu \left(\mu ^2+9\right)
{y^2_0}\right)+
\\& 5 \mu  \left(\mu ^2+3\right)  \frac{z^2}{L^2} \left(3 \left(5 \mu ^2-3\right) x^2_0-36 \mu
x_0 y_0-\left(\mu ^2-15\right) y^2_0 \right)+\\& 20 \mu ^2 \left(\mu ^2+3\right)
\frac{z^3}{L^3} (y_0-\mu x_0) \left(2 \mu y_0-\left(\mu ^2-3\right) x_0\right)-\\&
20 \mu ^3 \left(\mu ^2+3\right) \frac{z^4}{L^4} (y_0-\mu x_0)^2 \Big\}. \end{split}
\end{equation}

\begin{equation}\label{Ay2z}% it is correct
\begin{split}
y_2(z)=& -\frac{\mu/\rho }{270 (1+\mu ^2/3)^3  }\Big(1-\frac{z}{L}\Big)
\frac{z}{L}\times\\& \Big\{\left(7 \mu ^4-75 \mu ^2+360\right) x^2_0+6 \mu
\left(\mu ^2+75\right) x_0 y_0+3 \mu ^2 \left(5 \mu ^2+39\right) y^2_0 +\\& 2
\frac{z}{L} \left(\left(\mu ^6-4 \mu ^4+255 \mu ^2+180\right) x^2_0+\mu \left(\mu
^2+15\right) \left(13 \mu ^2+3\right) x_0 y_0+\mu ^2 \left(5 \mu ^4+36 \mu
^2-9\right) y^2_0\right)-\\& 14 \mu \left(\mu ^2+3\right)  \frac{z^2}{L^2} (y_0-\mu
x_0) \left(\left(15-4 \mu ^2\right) x_0+9 \mu y_0\right)+\\& 84 \mu ^2 \left(\mu
^2+3\right) \frac{z^3}{L^3} (y_0-\mu x_0)^2 \Big\}.
\end{split}
\end{equation}
In the leading, see Eq.~(\ref{Aactionkappa}), and next-to-leading order in
$1/\sqrt{\mathrm{SNR}}$ the action reads
\begin{equation}\label{AactinNLO}% it is correct
\begin{split}
&\frac{1}{Q}S[\Psi_{cl}(z)]=\frac{(1+4\mu^2/3)x^2_0-2\mu x_0
y_0+y^2_0}{QL(1+\mu^2/3)}+\frac{\mu/ \rho}{135  Q L \left(1+\mu ^2/3\right)^3
}\Big\{ \mu \left(4 \mu ^4+15 \mu ^2+225\right) x^3_0+ \\& { \left(23 \mu ^4+255 \mu
^2-90\right) x^2_0 y_0+\mu  \left(20 \mu ^4+117 \mu ^2-45\right) x_0 y^2_0-3 \left(5
\mu ^4+33 \mu ^2+30\right)   y^3_0} \Big\}+{\cal
O}\left(\frac{1}{{\mathrm{SNR}}}\right).
\end{split}
\end{equation}

\end{itemize}

%===============================
\subsection{The path-integral calculation.}\label{Appendix2}

To calculate the conditional probability density $P[Y|X]$ in
Eq.~(\ref{AQuasiclassProbabInitial}) one should find  the pre-exponent
path-integral, referred to as  the quantum corrections near the classical solution
$\Psi_{cl}(z)$,  in the leading and next-to-leading order in
$1/\sqrt{\mathrm{SNR}}$:
\begin{eqnarray}\label{Acuantumcorr0}
&&I_{QC}[\Psi_{cl}(z)]=\int \limits_{\tilde{\psi}(0)=0}^{{\psi}(L)=0}\!\!\!\!\!
{\cal D}{\psi} \,\exp
\left\{-\dfrac{S[\Psi_{cl}(z)+{\psi}(z)]-S[\Psi_{cl}(z)]}{Q}\right\}.
\end{eqnarray}
In what follows we are interested in the leading and next-to-leading order
corrections for the path-integral (\ref{AQuasiclassProbabInitial}). That is why we
retain only quadratic in $\psi$ terms in Eq.~(\ref{AdeltaS}). All these terms are
placed in the second line of Eq.~(\ref{AdeltaS}). As it will be demonstrated below
an extra power of $\psi$ results in an extra power of $\sqrt{QL}$. In the leading
and next-to-leading order calculation of the path-integral we should take into
account the first correction ($\varkappa_1(z) \propto \sqrt{QL}$) to the solution
$\Psi_{cl}(z)$, see Eqs.~(\ref{APsiclass}) and (\ref{Akappasolution}). Now we put
(\ref{APsiclass}) with $\varkappa_1(z)$ and
 $\psi(z)$ in the form $\psi(z)=u(z)\exp\left\{i \mu \frac{z}{L}+i
 \phi^{(X)}\right\}$ into the first line of Eq.~(\ref{AdeltaS}). In our approximation we obtain
\begin{eqnarray}\label{deltaS1}
&&{S[\Psi_{cl}(z)+{\psi}(z)]-S[\Psi_{cl}(z)]}= \int^{L}_{0}dz\Big\{ \left|
\partial_z u - i \frac{\mu}{L}(u+\bar{u}) \right|^2+  \nonumber \\&&
2 \frac{\mu}{L \rho} Im \left[2\Big(\partial_z \bar{u}+ i
\frac{\mu}{L}(u+\bar{u})\Big)\Big(u(\varkappa_1+\bar{\varkappa}_1)
+\bar{u}\varkappa_1 \Big)+\Big(\partial_z \bar{\varkappa}_1+i
\frac{\mu}{L}(\varkappa_1+\bar{\varkappa}_1) \Big)\Big(2|u|^2+u^2\Big) \right] +
\nonumber \\&& {\cal O}\left( \frac{\gamma}{L} u^2 \varkappa^2_1 \right)+ {\cal
O}\left( \frac{\gamma }{L}\rho u^2 \varkappa_2 \right)+{\cal O}\left( \gamma^2
\rho^2 u^4 \right)\Big\}.
\end{eqnarray}
We substitute this difference in the exponent in Eq.~(\ref{Acuantumcorr0}). Then we
expand the exponent at small $Q$ and obtain:
\begin{eqnarray}\label{Aexponent}
&&\exp \left\{-\dfrac{S[\Psi_{cl}(z)+{\psi}(z)]-S[\Psi_{cl}(z)]}{Q}\right\} =\exp
\Big\{-\frac{1}{Q}\int^{L}_0 dz \Big|\partial_z u - i \frac{\mu}{L}(u+\bar{u})
\Big|^2 \Big\}\Big\{1- \nonumber \\&& \frac{2\mu}{Q L \rho} Im \int^{L}_{0}dz
\left[2\Big(\partial_z \bar{u}+ i
\frac{\mu}{L}(u+\bar{u})\Big)\Big(u(\varkappa_1+\bar{\varkappa}_1)
+\bar{u}\varkappa_1 \Big)+\Big(\partial_z \bar{\varkappa}_1+i
\frac{\mu}{L}(\varkappa_1+\bar{\varkappa}_1) \Big)\Big(2|u|^2+u^2\Big) \right]
+\nonumber \\&& {\cal O}\left(\frac{QL}{\rho^2}\right)+{\cal O}\left(\gamma^2 QL^3
\rho^2 \right) \Big\}.
\end{eqnarray}
Here we imply that any extra power of $u$ or $\varkappa$ is suppressed by the
multiplicative parameter $\sqrt{QL}$, because at small $Q$ the main contribution to
the path-integral comes from  $u \sim \sqrt{QL}$. We substitute this expansion
(\ref{Aexponent}) into the path-integral (\ref{Acuantumcorr0}) and change the
variable from ${\psi}(z)$ to $u(z)$ and arrive at
\begin{eqnarray}\label{Acuantumcorr1}
&&I_{QC}[\Psi_{0}(z)]= \int \limits_{u(0)=0}^{u(L)=0}\!\!\!\!\! {\cal D}u \,\exp
\Big\{-\frac{1}{Q}\int^{L}_0 dz \Big|\partial_z u - i \frac{\mu}{L}(u+\bar{u})
\Big|^2 \Big\}\Big[1- \nonumber \\&& \frac{2\mu}{Q L \rho} Im \int^{L}_{0}dz
\left[2\Big(\partial_z \bar{u}+ i
\frac{\mu}{L}(u+\bar{u})\Big)\Big(u(\varkappa_1+\bar{\varkappa}_1)
+\bar{u}\varkappa_1 \Big)+\Big(\partial_z \bar{\varkappa}_1+i
\frac{\mu}{L}(\varkappa_1+\bar{\varkappa}_1) \Big)\Big(2|u|^2+u^2\Big) \right]
+\nonumber \\&& {\cal O}\left(\frac{QL}{\rho^2}\right)+{\cal O}\left(\gamma^2 QL^3
\rho^2 \right) \Big].
\end{eqnarray}
To calculate the leading and next-to-leading order contributions to
$I_{QC}[\Psi_{0}(z)]$ in $1/\mathrm{SNR}$ we should take the first and the second
terms in the square brackets in Eq.~(\ref{Acuantumcorr1}), respectively. We start
our consideration from the leading order. In this case we represent the
path-integral $I^{(0)}_{QC}[\Psi_{0}(z)]= \int_{u(0)=0}^{u(L)=0} {\cal D}u \,\exp
\Big\{-\frac{1}{Q}\int^{L}_0 dz \Big|\partial_z u - i \frac{\mu}{L}(u+\bar{u})
\Big|^2 \Big\}$ in the retarded discretization scheme:
\begin{eqnarray}\label{Acuantumcorr}
&&I^{(0)}_{QC}[\Psi_{0}(z)]= \int \limits_{u(0)=0}^{u(L)=0}\!\!\!\!\! {\cal D}u
\,\exp \Big\{-\frac{1}{Q}\int^{L}_0 dz \Big|\partial_z u - i
\frac{\mu}{L}(u+\bar{u}) \Big|^2 \Big\}= \nonumber \\&& \lim_{N \rightarrow \infty}
\left( \frac{N}{ \pi Q L}\right)^{N } \int^{\infty}_{-\infty}
\prod^{N-1}_{i=1}du^{(1)}_idu^{(2)} _i \exp\left\{- \frac{N}{Q
L}\sum^{N-1}_{i=0}\Big\{ (u^{(1)}_{i+1}-u^{(1)}_i)^2+(u^{(2)}_{i+1}-u^{(2)}_i -2
\frac{\mu}{N} u^{(1)}_i)^2\Big\}\right\},
\end{eqnarray}
where we use the measure (\ref{Ameasure}) and the notations $u(z_j)=u^{(1)}_j+i
u^{(2)}_j$, $z_i=\Delta \, i $, $\Delta=\frac{L}{N}$ and
$u^{(1)}_0=u^{(1)}_{N+1}=u^{(2)}_0=u^{(2)}_{N+1}=0$. The sequential integration over
$u^{(2)}_{N-1}$, $u^{(2)}_{N-2}$, $\ldots $, $u^{(2)}_{1}$ is trivial: \be \int d Y
\exp\left\{- \frac{(A-Y)^2}{2\tau_1}- \frac{(Y-B)^2}{2\tau_2}\right\}=\left( 2\pi
\frac{\tau_1 \tau_2 }{\tau_1+\tau_2}\right)^{1/2}\exp\left\{-
\frac{(A-B)^2}{2(\tau_1+\tau_2)}\right\}. \ee It leads to the remaining integral
(over $u^{(1)}_i$, $i=1, \ldots, N-1$) of the form
\begin{eqnarray}\label{Acuantumcorr2}
&&\lim_{N \rightarrow \infty} \left( \frac{N}{ \pi Q L}\right)^{N } \frac{(\pi Q
L/N)^{\frac{N-1}{2}}}{\sqrt{N}} \int^{\infty}_{-\infty} \prod^{N-1}_{i=1}du^{(1)}_i
 \exp\left\{- \frac{N}{Q L}\sum^{N-1}_{i,j=1}u^{(1)}_i M_{i,j}(\alpha)u^{(1)}_j\right\},
\end{eqnarray}
where we denote $\alpha=\frac{4}{N} \Big(\frac{\mu}{N}\Big)^2$, and the $(N-1)$ by
$(N-1)$  matrix $M(\alpha)$ has the following elements: $M_{i,i}=2+\alpha$,
$M_{i,i\pm 1}=-1+\alpha$, $i=1,\ldots, N-1$, $M_{i,j}=\alpha$, $j\neq i, j\neq i\pm
1$. It is straightforward to calculate the determinant of $M(\alpha)$ and hence to
perform the Gaussian integration over $u^{(1)}_i$
\begin{eqnarray} \label{detM}
\det[M(\alpha)]=N+\alpha \frac{N^2(N^2-1)}{12},
\end{eqnarray}
\begin{eqnarray} \label{leadingcontrib}
I^{(0)}_{QC}[\Psi_{0}(z)]=\frac{1}{\pi Q L \sqrt{1+\mu^2/3}}.
\end{eqnarray}

To calculate the next-to-leading  order contribution to the path-integral
(\ref{Acuantumcorr1}) we should take the second term in the square brackets in
Eq.~(\ref{Acuantumcorr1}). To find this correction we should calculate the integral
(the correlator):
\begin{eqnarray}\label{correlator0}
&& \langle  u^{(\alpha)} (z) u^{(\beta)} (z') \rangle \equiv
\frac{1}{I^{(0)}_{QC}[\Psi_{0}(z)]}\int \limits_{u(0)=0}^{u(L)=0}\!\!\!\!\! {\cal
D}u \,e^{-\frac{1}{Q}\int^{L}_0 dz \Big|\partial_z u - i \frac{\mu}{L}(u+\bar{u})
\Big|^2 } u^{(\alpha)} (z) u^{(\beta)} (z') =  Q L \,G^{\alpha,\,\beta}(z,z'),
\end{eqnarray}
where we have introduced the dimensionless Green matrix $G^{\alpha,\,\beta}(z,z')$,
$\alpha,\,\beta = 1,\,2$. The standard method for the Green matrix calculation is
the calculation of the generating functional \cite{Zinn-Justin}:
\begin{eqnarray}\label{generatingfunctional}
&&Z[J_1,J_2]=\int \limits_{u(0)=0}^{u(L)=0}\!\!\!\!\! {\cal D}u \,\exp
\Big\{-\frac{1}{Q}\int^{L}_0 dz \Big|\partial_z u - i \frac{\mu}{L}(u+\bar{u})
\Big|^2+ \int^{L}_0 dz  \left(J_1(z)u^{(1)} (z)+J_2(z)u^{(2)} (z) \right) \Big\},
\end{eqnarray}
then any correlator can be derived from the variation of the $Z[J_1,J_2]$ over
$J_{\alpha}$, for example
\begin{eqnarray}\label{correlator1}
&& \langle  u^{(\alpha)} (z) u^{(\beta)} (z') \rangle =  \frac{1}{Z[J_1,J_2]}
\frac{\delta Z[J_1,J_2]}{\delta J_{\alpha}(z) \delta
J_{\beta}(z')}\Big|_{J_1=0,\,J_2=0}.
\end{eqnarray}
The calculation of the generating functional can be performed  in the same way as
the calculation of the normalization integral (\ref{Acuantumcorr}): the integration
over $u^{(2)}_j$ followed by  the integration over $u^{(1)}_j$. The only new element
in the calculation of  the Gaussian integrals with the sources $J_{\alpha}$ is the
inverse matrix $M(\alpha)^{-1}_{i,j}$ for $M(\alpha)_{i,j}= \alpha+2
\delta_{i,j}-\delta_{i,j+1}-\delta_{i+1,j}$ defined herein above, see the text after
Eq.~(\ref{Acuantumcorr2}). The calculation is simple (after the observation that
$\det[M(\alpha)] M(\alpha)^{-1}_{i,j}$ is linear in $\alpha$), and we only set out
the result
\begin{eqnarray}\label{Minverse}
&& M(\alpha)^{-1}_{i,j}=N\left[ \frac{i}{N}\left(1- \frac{j}{N}\right)\theta(i \leq
j)+ \frac{j}{N}\left(1-\frac{i}{N}\right)\theta(i > j)-\frac{\alpha N^4}{4
\det[M(\alpha)]}\frac{i}{N}\left(1-\frac{i}{N}\right)\frac{j}{N}\left(1-\frac{j}{N}\right)
\right],
\end{eqnarray}
where $\det[M(\alpha)]$ is given by Eq.~  (\ref{detM}), and $\lim_{N \to
\infty}\left( \alpha N^4/(4\det[M(\alpha)])\right)=3 \mu^2/(3+\mu^2)$. We present
the result of the generating functional calculation in the form of a Green matrix
convolution with the sources $J_{\alpha}$:
\begin{eqnarray}\label{gfresult}
&&Z[J_1,J_2]= \frac{1}{\pi Q L
\sqrt{1+\mu^2/3}}\exp\left\{\frac{QL}{2}\int^{L}_{0}dz\int^{L}_{0}dz' J_{\alpha}(z)
G^{\alpha,\,\beta}(z,z') J_{\beta}(z')\right\} ,
\end{eqnarray}
where the Green matrix is Hermitian and it has the following elements:
\begin{eqnarray}\label{G11}
&&
G^{1,\,1}(z,z')=G^{1,\,1}(z',z)=\left\{\theta(z'-z)\frac{z}{2L}\left(1-\frac{z'}{L}\right)
-
\frac{3\mu^2}{4(3+\mu^2)}\left(1-\frac{z}{L}\right)\left(1-\frac{z'}{L}\right)\frac{z
z'}{L^2}\right\}+ \{z \leftrightarrow z' \},
\end{eqnarray}

\begin{eqnarray}\label{G12}
&& G^{1,\,2}(z,z')=G^{2,\,1}(z',z)=\frac{\mu}{2(3+\mu^2)}\Bigg\{
\theta(z-z')\frac{z'}{L}\left(1-\frac{z}{L}\right)\left(3\frac{z'}{L}-3\frac{z}{L}+
\frac{z'}{L}\mu^2\Big[1+ \frac{z}{L}\Big(2\frac{z'}{L}-3\Big) \Big] \right)+
\nonumber \\&&
\theta(z'-z)\frac{z}{L}\left(1-\frac{z'}{L}\right)\left(3\frac{z'}{L}-3\frac{z}{L}+\Big(\frac{z'}{L}-1\Big)
\mu^2 \Big[\frac{z}{L}+2\frac{z'}{L}\Big(\frac{z}{L}-1\Big) \Big]\right) \Bigg\},
\end{eqnarray}

\begin{eqnarray}\label{G22}
&& G^{2,\,2}(z,z')=G^{2,\,2}(z',z)= \Bigg\{\frac{\theta(z-z')}{6(3+\mu^2)}
\Big(1-\frac{z}{L}\Big)\frac{z'}{L}\Bigg[9+3 \mu^2 \Big(1+
\frac{z}{L}-2\frac{z^2}{L^2}+3\frac{z z'}{L^2}-2\frac{z'^2}{L^2} \Big)+ \nonumber
\\&& 2 \mu^4
\frac{z'}{L}\Big(\frac{z}{L}-1\Big)\Big(\frac{z'}{L}-3\frac{z}{L}+2 \frac{z
z'}{L^2}\Big) \Bigg]\Bigg\} + \{z \leftrightarrow z' \}.
\end{eqnarray}

The second way to obtain the expression for the correlator (\ref{correlator0}) and
Eqs.~(\ref{G11})-(\ref{G22}) reflects the fact that the Gaussian integral
(\ref{generatingfunctional}) is saturated in the vicinity of the saddle-point
solution of the equation of motion (i.e. the Euler-Lagrange equation for the action
in question) \cite{Feynman}. Thus to find it we should solve the set of equations
\begin{eqnarray}\label{GfEq}
&& \hat{K}_{\alpha,\gamma} G^{\gamma,\beta}(z,z')=
\frac{1}{L}\delta^{\beta}_{\alpha}\delta(z-z'),
\end{eqnarray}
where the matrix differentiation operator $\hat{K}$ for the functions
$u(z=0)=u(z=L)=0$ is defined as
\begin{eqnarray}
-\frac{1}{Q}\int^{L}_0 dz \Big|\partial_z u - i \frac{\mu}{L}(u+\bar{u})
\Big|^2=-\frac{1}{2 Q }\int^{L}_0 dz u^{(\alpha)}(z) \hat{K}_{\alpha,\beta}
u^{(\beta)}(z),
\end{eqnarray}
and it has the form
\begin{eqnarray}\label{Lab}
&& \hat{K}=2\begin{pmatrix}
-\partial^2_{z}+\frac{4\mu^2}{L^2}, & -2\frac{\mu}{L}\partial_{z} &  \\
2\frac{\mu}{L}\partial_{z}, & -\partial^2_{z}&
\end{pmatrix}.
\end{eqnarray}
The boundary conditions for equations (\ref{GfEq}) are as follows:
$G^{\alpha,\beta}(z=0,z')=G^{\alpha,\beta}(z=L,z')=0$. The problem has the
unambiguous solution (\ref{G11})-(\ref{G22}). Note that the homogeneous solution of
the Eq.~(\ref{GfEq}) is governed by the solutions of Eq.~ (\ref{Akappaequation})
obtained above.

Using the correlator (\ref{correlator0}) with (\ref{G11})-(\ref{G22}) one can easily
calculate the first correction presented in the second line of Eq.~
(\ref{Acuantumcorr1}). This term is proportional to $\varkappa_1(z) \propto
\sqrt{QL}$ hence delivering the first correction in $1/\sqrt{\mathrm{SNR}}$ to the
leading term (\ref{leadingcontrib}). The  subsequent integration of the elements
(\ref{G11})-(\ref{G22}) with the solution  (\ref{Akappasolution}) for
$\varkappa_1(z)$ is trivial, however the proper way to understand the discontinuous
derivatives of the Green matrix elements (\ref{G11})-(\ref{G22}) at the same point
$z'=z$ is the retarded scheme adopted in our approach \cite{Terekhov:2014}:
$\partial_z G^{\alpha,\beta}(z,z')|_{z'=z} \rightarrow \partial_z
G^{\alpha,\beta}(z+0,z')|_{z'=z}$. Finally we have
\begin{eqnarray}\label{Acuantumcorrfinal}
&&I_{QC}[\Psi_{0}(z)]=\frac{1}{\pi Q L \sqrt{1+\mu^2/3}}\Bigg[1 - \frac{\mu/\rho}{15
(1+\mu^2/3)^2}\left(\mu (15+\mu^2)x_0-2(5-\mu^2/3)y_0\right) + \nonumber
\\&&{\cal O}\left(\frac{QL}{\rho^2}\right) +{\cal O}\left(\gamma^2 QL^3 \rho^2
\right)\Bigg].
\end{eqnarray}
The error estimation ${\cal O}\left({QL}/{\rho^2}\right)={\cal
O}\left(1/\mathrm{SNR}\right)$ comes from taking into account the next term in the
$\Psi_{cl}$ expansion of the complete action expression (\ref{AdeltaS}). The
estimation ${\cal O}\left(\gamma^2 QL^3 \rho^2 \right)$ appears as the estimation of
the contribution of the nonlinear biquadratic terms, see Eq.~(\ref{deltaS1}),
originating from the third line in the action (\ref{AdeltaS}). Indeed, it is obvious
from the expression (\ref{correlator0}) that any extra power of the field $u(z)$
results in an extra factor $\sqrt{QL}$. Substituting the leading order expression
for $\Psi_{cl}$ in the third line of Eq.~(\ref{AdeltaS}) for these terms we arrive
at the estimation ${\cal O}\left(\gamma^2 QL^3 \rho^2 \right)$. That is why the
formal parameter of the perturbation theory for the nonlinear terms is  $QL \gamma^2
L^2 \rho^2$. Note the quantity $QL \gamma^2  L^2 \rho^2 \equiv N \gamma^2 L^2 P$ is
the very parameter determining the upper limit of the intermediate power region
(\ref{region}).

Finally, from Eq.~ (\ref{AactinNLO}) for the exponent factor and from
Eq.~(\ref{Acuantumcorrfinal})  for the pre-exponent factor  we arrive at the
expression
\begin{eqnarray}\label{APYXapprox}
&&\!\!\!P[Y|X]=\dfrac{\exp\left\{- \dfrac{(1+4\mu^2/3)x^2_0-2\mu x_0 y_0+y^2_0}{Q L
(1+\mu^2/3)}\right\}} {\pi Q L \sqrt{1+\mu^2/3}}\Bigg(1- \frac{\mu/\rho}{15
(1+\mu^2/3)^2}\left(\mu (15+\mu^2)x_0-2(5-\mu^2/3)y_0\right)- \nonumber \\&&
\frac{\mu/ \rho}{135 Q L \left(1+\mu ^2/3\right)^3 }\Big\{ \mu \left(4 \mu ^4+15 \mu
^2+225\right) x^3_0+ \left(23 \mu ^4+255 \mu ^2-90\right) x^2_0 y_0+  \mu \left(20
\mu ^4+117 \mu ^2-45\right) x_0 y^2_0- \nonumber \\&& 3 \left(5 \mu ^4+33 \mu
^2+30\right) y^3_0 \Big\}+ {\cal O}\Big(\frac{QL}{\rho^2}\Big)+ {\cal
O}\left(\gamma^2 QL^3 \rho^2 \right) \Bigg).
\end{eqnarray}
Now it is easy to show, that the normalization condition
\begin{eqnarray}\label{APYXnormalization}
&&\int DY P[Y|X]=1
\end{eqnarray}
is fulfilled.

%===============================
\subsection{Calculation of $P_{out}[Y]$.}\label{Appendix3}

Let us consider the integral $P_{out}[Y]=\int {\cal D}X P_X[X]P[Y|X]$. In our case
the measure ${\cal D}X = dxdy$, where $x=Re\{X\}$, $y=Im\{X\}$, so we should
consider the integral:
\begin{eqnarray}\label{LaplaceInt1}
\int\limits_{-\infty}^\infty dxdy P_X[x,y]P[Y|X]\,.
\end{eqnarray}
In the integral the scale of variation of the function $P_X[x,y]$ is $P\gg QL$. The
scale of variation of the function $P[Y|X]$ is $QL$, and this function has the form
Eq. (\ref{PYXapprox}), therefore we can use Laplace's method. To demonstrate that
one can see that the function $P[Y|X]$ depends on $|X|$,
$x_0=Re\{\bar{X}(Ye^{-i\mu}-X)/|X|\}$, $y_0=Im\{\bar{X}(Ye^{-i\mu}-X)/|X|\}$, and
reaches the maximal value at the point $x_0=y_0=0$. Let us change the integration
variables $x,\,y$ to $ \eta_1,\,\eta_2$, where $\eta=\eta_1+i\eta_2=(X
e^{i{\mu}}-Y)e^{-i\phi^{(Y)}}$. Here $\phi^{(Y)}$ is the phase of the $Y$. The
inverse transformation reads $X=(\eta+|Y|)e^{-i \gamma L |\eta+|Y||^2+i\phi^{(Y)}}$.
In the new variables the function $P[Y|X]$ reaches maximum at the point
$\eta_1=\eta_2=0$. The integral (\ref{LaplaceInt1}) takes the following form
\begin{eqnarray}\label{LaplaceInt2}
\int\limits_{-\infty}^\infty d\eta_1d\eta_2 P_X\left[Re\left\{(\eta+|Y|)e^{-i\gamma L
|\eta+|Y||^2+i\phi^{(Y)}}\right\},Im\left\{(\eta+|Y|)e^{-i\gamma L
|\eta+|Y||^2+i\phi^{(Y)}}\right\}\right]P[Y|X]\,,
\end{eqnarray}
here we have used the fact that the Jacobian determinant for the variables
transformation is equal to unity. Since $P[Y|X]$ reaches its maximum at the point
$\eta=0$ we can expand the  functions $P_X[X]$ and $P[Y|X]$ in the vicinity of the
point:
\begin{eqnarray}
&&\!\!\!\!\!\!\!\!\!\!\!\!\!\!\!P_X[Re\{(\eta+|Y|)e^{-i\gamma L
|\eta+|Y||^2+i\phi^{(Y)}}\}, Im\{(\eta+|Y|)e^{-i\gamma L
|\eta+|Y||^2+i\phi^{(Y)}}\}] \approx \nonumber \\&& \left(
P_X[Re\{Ye^{-i\tilde{\mu}}\}, Im\{Ye^{-i\tilde{\mu}}\}]+\mathrm{terms\,\,
proportional\,\, to\,\, }\,\,\eta+\ldots \right)\,,
\label{expansionPX}\\
&&\!\!\!\!\!\!\!\!\!\!\!\!\!\!\!P[Y|X]\approx\frac{1}{\pi
QL\sqrt{1+\tilde{\mu}^2/3}}\exp\left\{-\frac{(1+4\tilde{\mu}^2/3)\eta_1^2-2\tilde{\mu}\eta_1\eta_2+
\eta_2^2}{QL(1+\tilde{\mu}^2/3)}\right\}\left(1+\mathrm{terms\,\, proportional\,\,
to\,\, }\,\,\eta \,\,\,\,\mathrm{and } \frac{\eta^3}{QL}
\right)\label{expansionPYX}\,,
\end{eqnarray}
where we have used the fact that in the vicinity of the point $\eta=0$ we have
$x_0=-\eta_1$ and $y_0=-\eta_2$ up to higher powers of $\eta$. In Eqs.~
(\ref{expansionPX}) and (\ref{expansionPYX}) we have the parameter
$\tilde{\mu}=\gamma L |Y|^2$.

One can see that at large $\tilde{\mu}$ the exponent contains three different terms:
\begin{eqnarray}
\frac{(1+4\tilde{\mu}^2/3)\eta_1^2-
2\tilde{\mu}\eta_1\eta_2+\eta_2^2}{QL(1+\tilde{\mu}^2/3)}\approx
\frac{4\eta_1^2}{QL}-6\frac{\eta_1\eta_2}{QL\tilde{\mu}}+\frac{3\eta_2^2-9\eta_1^2}{QL\tilde{\mu}^2}\,.
\end{eqnarray}
Therefore to use Laplace's method we  have to transform our quadratic form
\begin{eqnarray}
(\eta_1,\eta_2)A(\eta_1,\eta_2)^T=\frac{(1+4\tilde{\mu}^2/3)\eta_1^2-2\tilde{\mu}\eta_1\eta_2+\eta_2^2}{QL(1+\tilde{\mu}^2/3)}
\end{eqnarray}
to the canonical form. The matrix of quadratic form is:
\begin{eqnarray}
A=\frac{1}{QL\left(1+\tilde{\mu}^2/3\right)}\left(
\begin{array}{cc}
1+\dfrac{4\tilde{\mu}^2}{3}&-\tilde{\mu}\\
-\tilde{\mu}&1
\end{array}
\right).
\end{eqnarray}
The eigenvalues of the matrix $A$  are
\begin{eqnarray}
\lambda_1&=&\frac{1}{QL}\left(1+\tilde{\mu}\frac{\tilde{\mu}+\sqrt{9+4\tilde{\mu}^2}}{3+\tilde{\mu}^2}\right)\,,\\
\lambda_2&=&\frac{1}{QL}\left(1+\tilde{\mu}\frac{\tilde{\mu}-\sqrt{9+4\tilde{\mu}^2}}{3+\tilde{\mu}^2}\right)\,.
\end{eqnarray}
One can see that $\lambda_{1,2}>0$, and at large $\tilde{\mu}$ they have the form:
\begin{eqnarray}
\lambda_1\approx \frac{4}{QL}\,,\quad\lambda_2\approx\frac{3}{4QL\tilde{\mu}^2}.
\end{eqnarray}
Therefore at large $\mu\approx\tilde{\mu}$ there are two  parameters in the Laplace
integral, one parameter is $1/QL$, the other is $1/(QL\tilde{\mu}^2)$. To use
Laplace's method for the integral Eq. (\ref{LaplaceInt1}) we have to impose two
conditions $P\gg QL$, and $P\gg QL\tilde{\mu}^2$. These conditions lead to the two
dimensionless parameters for Laplace's method
\begin{eqnarray}
&&\mathrm{SNR}\gg1\,,\\
&&(\gamma^2QL^3 P)^{-1}\gg 1.
\end{eqnarray}
To calculate the integral Eq.~(\ref{LaplaceInt2}) in  the leading order in the
parameters $1/\mathrm{SNR}$ and $(\gamma^2QL^3 P)$ we substitute the first term of
the expansion Eq. (\ref{expansionPX}) and the first term in the brackets of the
expansion Eq. (\ref{expansionPYX}) to the integral Eq.~(\ref{LaplaceInt1}). After
straightforward calculation we obtain:
\begin{eqnarray}
P_X\left[Re\left\{Ye^{-i\gamma L|Y|^2}\right\}, Im\left\{Ye^{-i\gamma L|Y|^2}\right\}\right]
\int\limits_{-\infty}^\infty d\eta_1d\eta_2 P[Y|X]\approx P_X\left[Ye^{-i\gamma
L|Y|^2}\right]\,.
\end{eqnarray}
To calculate corrections to the integral in  parameters $1/\mathrm{SNR}$ and
$\gamma^2QL^3 P$ we should take terms which are  proportional to $\eta$ and $\eta^3$
in the product of expansions Eqs. (\ref{expansionPX}) and (\ref{expansionPYX}).
Formally  the first correction to the integral should be of order of
$1/\sqrt{\mathrm{SNR}}$ and $\sqrt{\gamma^2QL^3 P}$, but it is zero due to the
symmetry $\eta\to-\eta$ (the exponent contains only even combination of $\eta$).
Therefore we can estimate the order of the first nonzero corrections as  ${\cal
O}(1/\mathrm{SNR})$ and ${\cal O}(\gamma^2 QL^3 P)$.  Therefore the result for the
integral Eq. (\ref{LaplaceInt1}) can be written as:
\begin{eqnarray}\label{LaplaceInt3}
\int\limits_{-\infty}^\infty dxdy P_X[x,y]P[Y|X]=P_X\left[Ye^{-i\gamma
L|Y|^2}\right]\left(1+{\cal O}\left(1/\mathrm{SNR}\right)+{\cal O}\left(\gamma^2QL^3
P\right)\right)\,.
\end{eqnarray}

%===========================================Bibliography

\end{document}